\pdfoutput=1
\documentclass[useAMS,usenatbib,usegraphicx]{mn2e}
\usepackage{apjfonts}
\usepackage{amssymb}
\usepackage{amsmath}
\usepackage{ctable}
\usepackage{fixltx2e} %% forces figure order to remain fixed (otherwise figure*'s can move out-of-order)

\newcommand{\Msun}{{\rm M_{\odot}}}
\newcommand{\nc}{{n_{\rm th}}}

\newcommand{\muv}{M_{\rm UV}}
\newcommand{\fesc}{f_{\rm esc}}

\newcommand{\Rvir}{R_{\rm vir}}
\newcommand{\Qint}{Q_{\rm int}}
\newcommand{\Qesc}{Q_{\rm esc}}
\newcommand{\pc}{{\rm pc}}
\newcommand{\kpc}{{\rm kpc}}
\newcommand{\Mpc}{{\rm Mpc}}
\newcommand{\cm}{{\rm cm}}

\newcommand{\hinv}{h^{-1}}

\title[Escape Fractions from FIRE Galaxies]
{The Difficulty of Getting High Escape Fractions of Ionizing Photons from High-redshift Galaxies: a View from the FIRE Cosmological Simulations}

\author[Ma et al.]{
  \parbox[t]{1.0\textwidth}{ 
   Xiangcheng Ma,$^1$\thanks{E-mail: xchma@caltech.edu}
   Daniel Kasen,$^{2,3}$
   Philip F. Hopkins,$^1$
   Claude-Andr{\'e} Faucher-Gigu{\`e}re,$^4$
   Eliot Quataert,$^2$
   Du{\v s}an Kere{\v s}$^5$ \&
   Norman Murray$^6$\thanks{Canada Research Chair in Astrophysics.}
   } 
  \vspace{5pt} \\
  $^1$TAPIR, MC 350-17, California Institute of Technology, Pasadena, CA 91125, USA \\ 
  $^2$Department of Astronomy and Theoretical Astrophysics Center, University of California Berkeley, Berkeley, CA 94720 \\
  $^3$ Lawrence Berkeley National Laboratory, 1 Cyclotron Road, Berkeley, CA 94720 \\
  $^4$Department of Physics and Astronomy and CIERA, Northwestern University, 2145 Sheridan Road, Evanston, IL 60208, USA \\
  $^5$Department of Physics, Center for Astrophysics and Space Sciences, University of California at San Diego, 9500 Gilman Drive, La Jolla, CA 92093 \\
  $^6$Canadian Institute for Theoretical Astrophysics, 60 St George Street, University of Toronto, ON M5S 3H8, Canada
}

\pagerange{\pageref{firstpage}--\pageref{lastpage}}
\date{Draft version \today}

\begin{document}
\maketitle
\label{firstpage}

\begin{abstract}
We present a series of high-resolution (20--2000 $\Msun$, 0.1--4 pc) cosmological zoom-in simulations at $z\gtrsim6$ from the Feedback In Realistic Environment (FIRE) project. These simulations cover halo masses $10^9$--$10^{11}~\Msun$ and rest-frame ultraviolet magnitude $\muv=-9$ to $-19$. These simulations include explicit models of the multi-phase ISM, star formation, and stellar feedback, which produce reasonable galaxy properties at $z=0$--6. We post-process the snapshots with a radiative transfer code to evaluate the escape fraction ($\fesc$) of hydrogen ionizing photons. We find that the instantaneous $\fesc$ has large time variability (0.01\%--20\%), while the time-averaged $\fesc$ over long time-scales generally remains $\lesssim5\%$, considerably lower than the estimate in many reionization models. We find no strong dependence of $\fesc$ on galaxy mass or redshift. In our simulations, the intrinsic ionizing photon budgets are dominated by stellar populations younger than 3 Myr, which tend to be buried in dense birth clouds. The escaping photons mostly come from populations between 3--10 Myr, whose birth clouds have been largely cleared by stellar feedback. However, these populations only contribute a small fraction of intrinsic ionizing photon budgets according to standard stellar population models. We show that $\fesc$ can be boosted to high values, if stellar populations older than 3 Myr produce more ionizing photons than standard stellar population models (as motivated by, e.g., models including binaries). By contrast, runaway stars with velocities suggested by observations can enhance $\fesc$ by only a small fraction. We show that ``sub-grid'' star formation models, which do not explicitly resolve star formation in dense clouds with $n\gg1~\cm^{-3}$, will dramatically over-predict $\fesc$. 
\end{abstract}


\section{Introduction}
\label{intro}
Star-forming galaxies at high redshifts are thought to be the dominant source of hydrogen reionization \citep[e.g.][]{madau.99.uvb,fg.08.uvb,haardt.12.uvb}. Therefore, the escape fraction of hydrogen ionizing photons ($\fesc$)  from these galaxies is an important, yet poorly constrained, parameter in understanding the reionization history. 

Models of cosmic reionization are usually derived from the galaxy ultraviolet luminosity function \citep[UVLF; e.g.][]{bouwens.11.uvlf,mclure.13.uvlf}, Thomson scattering optical depths inferred from Cosmic Microwave Background (CMB) measurements \citep{hinshaw.13.wmap, planck.13}, Ly$\alpha$ forest transmission \citep[e.g.][]{fan.06.gp}. They often require high $\fesc$ in order to match the ionization state of the intergalactic medium (IGM) by $z=6$ \citep[e.g.][]{ouchi.09,kuhlen.fg.12,finkelstein.12.candel,robertson.13.udf12}. For example, \citet{finkelstein.12.candel} and \citet{robertson.13.udf12} suggested $\fesc>13\%$ and $\fesc>20\%$, respectively, assuming all the ionization photons are contributed by galaxies brighter than $\muv=-13$. However, such constraints on $\fesc$ are always entangled with the uncertainties at the faint end of UVLF, since low-mass galaxies can play a dominant role in providing ionizing photons due to their dramatically increasing numbers. For example, \citet{finkelstein.12.candel} derived that reionization requires a much higher escape fraction $\fesc>34\%$ if one only accounts for the contribution of galaxies brighter than $\muv=-18$. Also, \citet{kuhlen.fg.12} showed that even applying a cut off on UV magnitude at $\muv=-13$, the required escape fraction at $z=6$ varies from 6\%--30\% when changing the faint-end slope of UVLF within observational uncertainties. Furthermore, it is also not clear how $\fesc$ depends on galaxy mass and evolves with redshift, which makes the problem more complicated.

Therefore, independent constraints on $\fesc$ are necessary to disentangle these degeneracies. Star-forming galaxies at lower redshifts should provide important insights into their high-redshift counterparts. In the literature, high escape fractions from 10\% up to unity have been reported in various samples of Lyman break galaxies (LBGs) and Ly$\alpha$ emitters (LAEs) around $z\sim3$ \citep[e.g.][]{steidel.01,shapley.06,vanzella.12,nestor.13}. These measurements are based on detection of {\it rest-frame} Lyman continuum (LyC) emission from either individual galaxies or stacked samples, so the exact value of $\fesc$ depends on uncertain dust and IGM attenuation correction. Similar observations at lower redshifts always show surprisingly low escape fractions. In the local universe, the only two galaxies which have confirmed LyC detection suggest $\fesc$ to be only $\sim2\%$--$3\%$ \citep{leitet.11,leitet.13}. At $z\sim1$, stacked samples have been used to derive upper limits as low as $\fesc<1\%$--$2\%$ \citep[e.g.][]{cowie.09,siana.10,bridge.10}. Even at $z\sim3$, low escape fractions ($<5\%$) have also been reported in some galaxy samples \citep[e.g.][]{iwata.09,boutsia.11}. Recent careful studies have revealed that a considerable fraction of specious LyC detection at $z\sim3$ is due to contamination from foreground sources (\citealt{vanzella.10}; for a very recent study, see \citealt{siana.15.lyc}), which could at least partly account for the apparent contradiction between these observations. Nevertheless, given the large uncertainty in these studies, no convincing conclusion can be reached so far from current observations.

Previous numerical simulations of galaxy formation also predict a broad range of $\fesc$, and even contradictory trends of the dependence of $\fesc$ on halo mass and redshift. For example, \citet{razoumov.10} found $\fesc$ decreases from unity to a few precent with increasing halo mass from $10^{7.8}$--$10^{11.5}~\Msun$. Similarly, \citet{yajima.11} also found their $\fesc$ decreases from 40\% at halo mass $10^9~\Msun$ to 7\% at halo mass $10^{11}~\Msun$. On the other hand, \citet{gnedin.08} found increasing $\fesc$ with halo mass in $10^{10}$--$10^{12}~\Msun$. They also reported significantly lower escape fraction of $1\%-3\%$ for the most massive galaxies in their simulations and $<0.1\%$ for the smaller ones. \citet{razoumov.10} also found $\fesc$ decreases from $z=4$--10 at fixed halo mass, while \citet{yajima.11} found no dependence of $\fesc$ on redshift. At lower masses, \citet{wise.cen.09} found $\fesc\sim5\%$--$40\%$ and $\fesc\sim25\%$--$80\%$ by invoking a normal initial mass function (IMF) and a top-heavy IMF, respectively, for galaxies of halo mass in $10^{6.5}$--$10^{9.5}~\Msun$; whereas \citet{paardekooper.11} reported lower escape fraction of $10^{-5}$--0.1 in idealized simulations of galaxy masses $10^8$--$10^9~\Msun$. 

Most of the intrinsic ionizing photons are produced by massive stars of masses in 10--100 $\Msun$, which are originally born in giant molecular clouds (GMCs). The majority of the ionizing photons are instantaneously absorbed by the dense gas in the GMCs and generate H II regions. These ``birth clouds'' must be disrupted and dispersed by radiation pressure, photoionization, H II thermal pressure, and supernovae before a considerable fraction of ionizing photons are able to escape \citep[e.g.][]{murray.10.radiation,kim.13.escape,paardekooper.15}. Therefore, to study the escape fraction of ionizing photons using simulations, one must resolve the multi-phase structure of the interstellar medium (ISM) and ``correctly'' describe star formation and stellar feedback. Many previous simulations adopt very approximate or ``sub-grid'' ISM and feedback model, which can lead to many differences between those studies. Recent studies have noted the importance of resolving the ISM structure around the stars and started to adopt more detailed treatments of the ISM and stellar feedback physics \citep{kim.13.escape,kimm.cen.14,wise.14,paardekooper.13,paardekooper.15}. For example, \citet{wise.14} performed radiative hydrodynamical simulations with state-of-art ISM physics and chemistry, star formation, and stellar feedback models and found $\fesc$ drops from 50\% to 5\% with increasing halo mass in $10^7$--$10^{8.5}~\Msun$ at $z>7$. They conclude that more massive galaxies are not likely to have high escape fractions, but are unable to simulate more massive systems. \citet{kimm.cen.14} explored more physically motivated models of supernovae (SN) feedback and found average escape fraction of $\sim11\%$ for galaxies in $10^8$--$10^{10.5}~\Msun$. \citet{paardekooper.15} argued that the dense gas within 10 pc from young stars provides the main constraint on the escape fraction. They found in their simulation that about 70\% of the galaxies of halo mass above $10^8~\Msun$ have escape fraction below 1\%. But none of these simulations has been run to $z=0$ to confirm that the models for star formation, feedback, and the ISM produce reasonable results in comparison to observations.

The Feedback in Realistic Environment (FIRE) project\footnote{FIRE project website: http://fire.northwestern.edu} \citep{hopkins.14.fire} is a series of cosmological zoom-in simulations that are able to follow galaxy merger histories, interaction of galaxies with IGM, and many other processes. The simulations include a full set of realistic models of the multi-phase ISM, star formation, and stellar feedback. The first series of FIRE simulations run down to $z=0$ reproduce reasonable star formation histories, the stellar mass-halo mass relation, the Kennicutt--Schmidt law, and the star-forming main sequence, for a broad range of galaxy masses ($M_{\ast}=10^4$--$10^{11}~\Msun$) from $z=0$--6 \citep{hopkins.14.fire}. Cosmological simulations with the FIRE stellar feedback physics self-consistently generate galactic winds with velocities and mass loading factors broadly consistent with observational requirements \citep{muratov.15.outflow} and are in good agreement with the observed covering fractions of neutral hydrogen in the halos of $z=2$--3 LBGs \citep{cafg.14.fire}. In previous studies of isolated galaxy simulations, these models have also been shown to reproduce many small scale observations, including the observed multi-phase ISM structure, density distribution of GMCs, GMC lifetimes and star formation efficiencies, and the observed Larson's law scalings between cloud sizes and structural properties, from scales $<1~\pc$ to $>\kpc$ \citep[e.g.][]{hopkins.11.fb,hopkins.12.fb}. A realistic model with these properties is {\it necessary} to study the production and propagation of ionizing photons inside a galaxy. 

In this work, we present a separate set of cosmological simulations at $z>6$, performed with the same method and models at extremely high resolution (particle masses 20--2000 $\Msun$, smoothing lengths 0.1--4 pc). These simulations cover galaxy halo masses $10^9$--$10^{11}~\Msun$ and rest-frame ultraviolet magnitudes $\muv=-9$ to $-19$ at $z=6$. We then evaluate the escape fraction of ionizing photons with Monte Carlo radiative transfer calculations. We describe the simulations and present the properties of our galaxies in Section \ref{sec:simulations} and \ref{sec:galaxy}. In Section \ref{sec:escfrac}, we describe the Monte Carlo radiative transfer code and compile our main results on the escape fractions and there dependence on galaxy mass and cosmic time. In Section \ref{sec:discussion}, we show how the UV background and star formation prescriptions affect our results. We also discuss the effects of runaway stars and extra ionizing photon budgets contributed by intermediate-age stellar populations, as motivated by recent observations and stellar models. We summarize and conclude in Section \ref{sec:conclusion}.

% Table 1
\begin{table*}
\begin{center}
\caption{The simulations.} 
\label{table:simlist}
\begin{tabular}{lcccccccccc}
\hline\hline
Name & Boxsize & $M_{\rm vir}$ & $m_b$ & $\epsilon_b$ & $m_{\rm dm}$ & $\epsilon_{\rm dm}$ & $\nc$ & $M_{\ast}$ & $\rm M_{UV}$ & Resolution \\
 & ($\hinv~\Mpc$) & ($\Msun$) & ($\Msun$) & ($\pc$) & ($\Msun$) & ($\pc$) & ($\cm^{-3}$) & ($\hinv~\Msun$) & ($AB~mag$) & \\
\hline
{\bf z5m09} & 1 & 7.6e8 & 16.8 & 0.14 & 81.9 & 5.6 & 100 & 3.1e5 & -10.1 & HR \\
{\bf z5m10}  & 4 & 1.3e10 & 131.6 & 0.4 & 655.6 & 7 & 100 & 2.7e7 & -14.8 & HR \\
{\bf z5m10mr} & 4 & 1.5e10 & 1.1e3 & 1.9 & 5.2e3 & 14 & 100 & 5.0e7 & -17.5 & MR  \\
{\bf z5m10e}$^a$ & 4 & 1.3e10 & 1.1e3 & 1.9 & 5.2e3 & 14 & 1 & 2.4e7 & -16.1 & MR \\
{\bf z5m10h} & 4 & 1.3e10 & 1.1e3 & 1.9 & 5.2e3 & 14 & 1000 & 6.6e7 & -16.4 & MR \\
{\bf z5m11} & 10 & 5.6e10 & 2.1e3 & 4.2 & 1.0e4 & 14 & 100 & 2.0e8 & -18.5 & MR \\
\hline\hline
\multicolumn{11}{p{0.8\linewidth}}{Initial conditions of our simulations and simulated galaxy properties at $z=6$:} \\
\multicolumn{11}{p{0.8\linewidth}}{(1) Name: Simulation designation.} \\
\multicolumn{11}{p{0.8\linewidth}}{(2) Boxsize: Zoom-in box size of our simulations.} \\
\multicolumn{11}{p{0.8\linewidth}}{(3) $M_{\rm vir}$: Halo mass of the primary galaxy at $z=6$.} \\
\multicolumn{11}{p{0.8\linewidth}}{(4) $m_b$: Initial baryonic (gas and star) particle mass in the high-resolution region.} \\
\multicolumn{11}{p{0.8\linewidth}}{(5) $\epsilon_b$: Minimum baryonic force softening (minimum SPH smoothing lengths are comparable or smaller). Force softening is adaptive (mass resolution is fixed).} \\
\multicolumn{11}{p{0.8\linewidth}}{(6) $m_{\rm dm}$: Dark matter particle mass in the high-resolution regions.} \\
\multicolumn{11}{p{0.8\linewidth}}{(7) $\epsilon_{\rm dm}$: Minimum dark matter force softening (fixed in physical units at all redshifts).} \\
\multicolumn{11}{p{0.8\linewidth}}{(8) $\nc$: Density threshold of star formation (see Section \ref{sec:simulations}).} \\
\multicolumn{11}{p{0.8\linewidth}}{(9) $M_{\ast}$: Stellar mass of the primary galaxy at $z=6$.} \\
\multicolumn{11}{p{0.8\linewidth}}{(10) $\rm M_{UV}$: Galaxy UV magnitude (absolute AB magnitude at 1500 \AA).} \\
\multicolumn{11}{p{0.8\linewidth}}{(11) Resolution: Whether a simulation is of ultra-high resolution (HR) or of medium resolution (MR).} \\
\multicolumn{11}{p{0.8\linewidth}}{Note:} \\
\multicolumn{11}{p{0.8\linewidth}}{$^a$ This simulation is intentionally designed to mimic ``sub-grid'' star formation models that are usually adopted in low-resolution cosmological simulations. We not only lower the star formation density threshold to $\nc=1~\cm^{-3}$, but also allow star formation at 2\% efficiency per free-fall time if gas reaches $\nc$ but is not self-gravitating.} \\
\hline\hline
\end{tabular}
\end{center}
\end{table*}

% Figure 1
\begin{figure*}
\centering
\begin{tabular}{ccc}
  \includegraphics[width=0.3\textwidth]{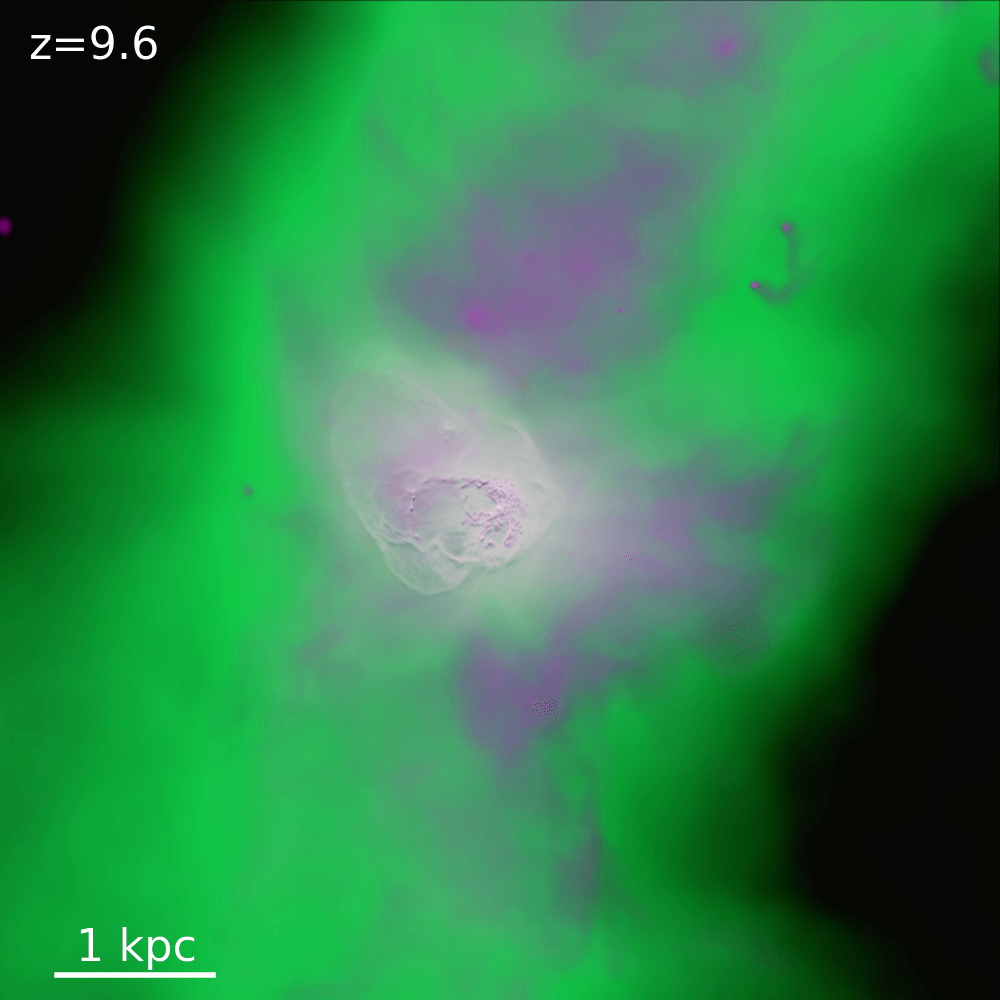} & 
  \includegraphics[width=0.3\textwidth]{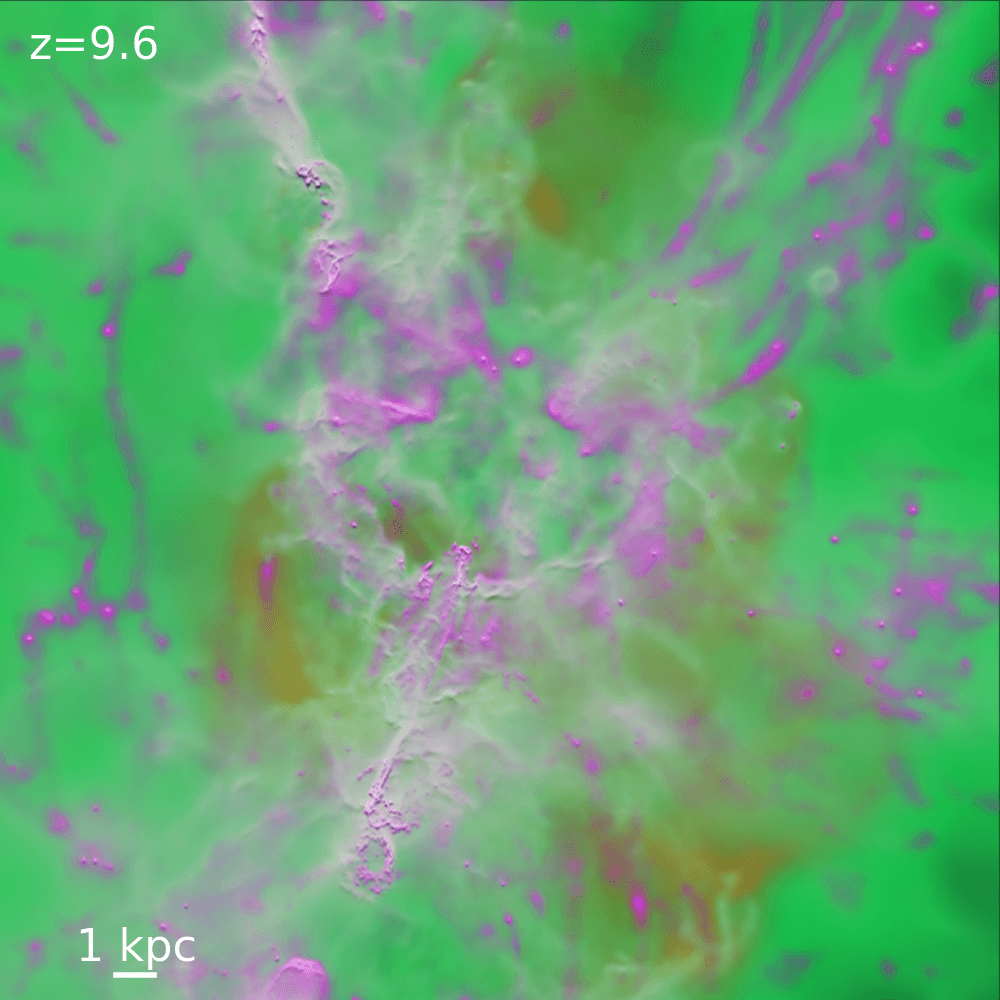} &
  \includegraphics[width=0.3\textwidth]{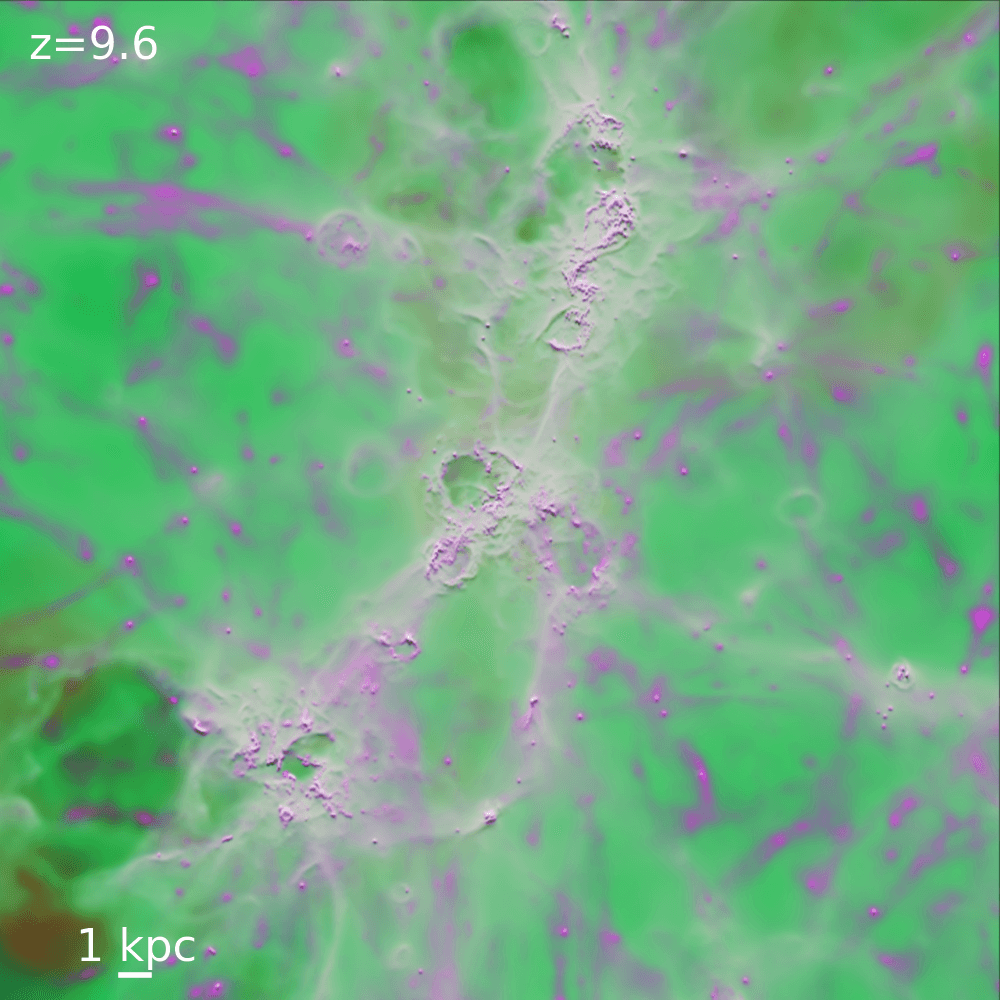} \\
  \includegraphics[width=0.3\textwidth]{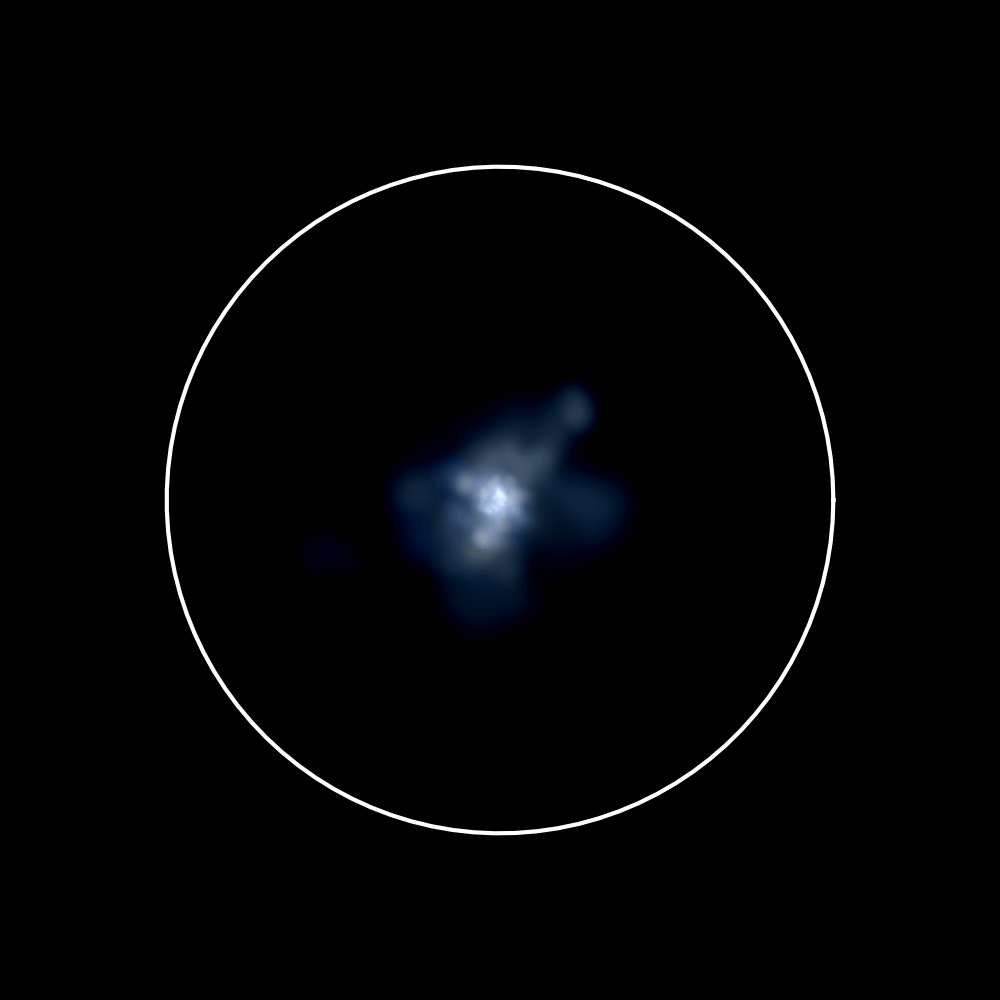} &
  \includegraphics[width=0.3\textwidth]{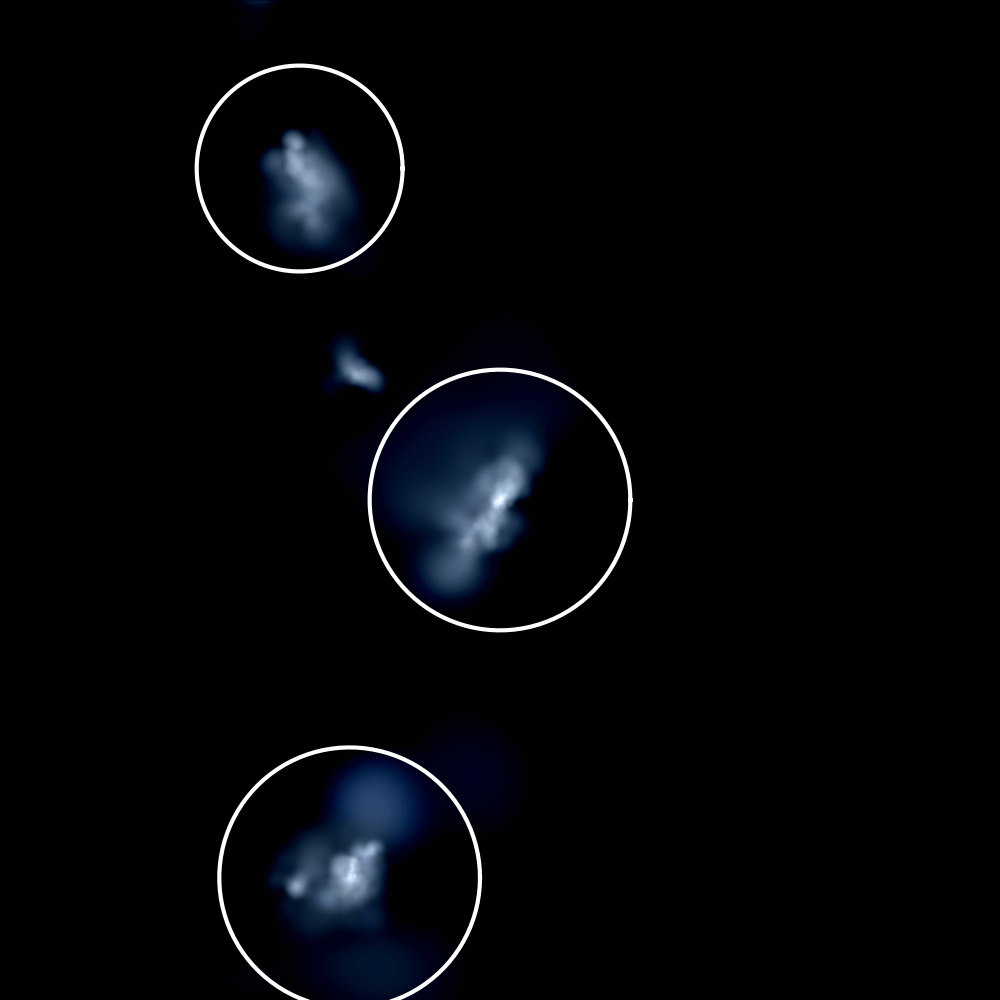} &
  \includegraphics[width=0.3\textwidth]{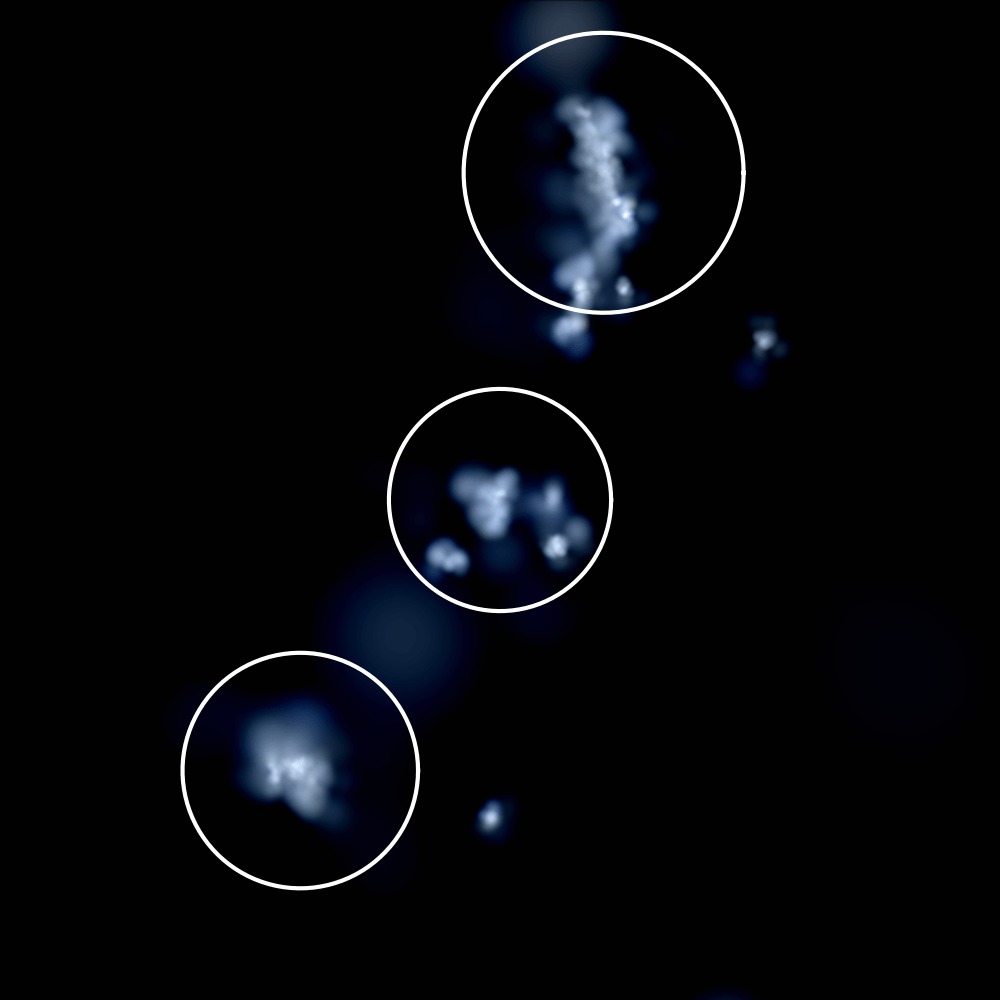} \\
  \includegraphics[width=0.3\textwidth]{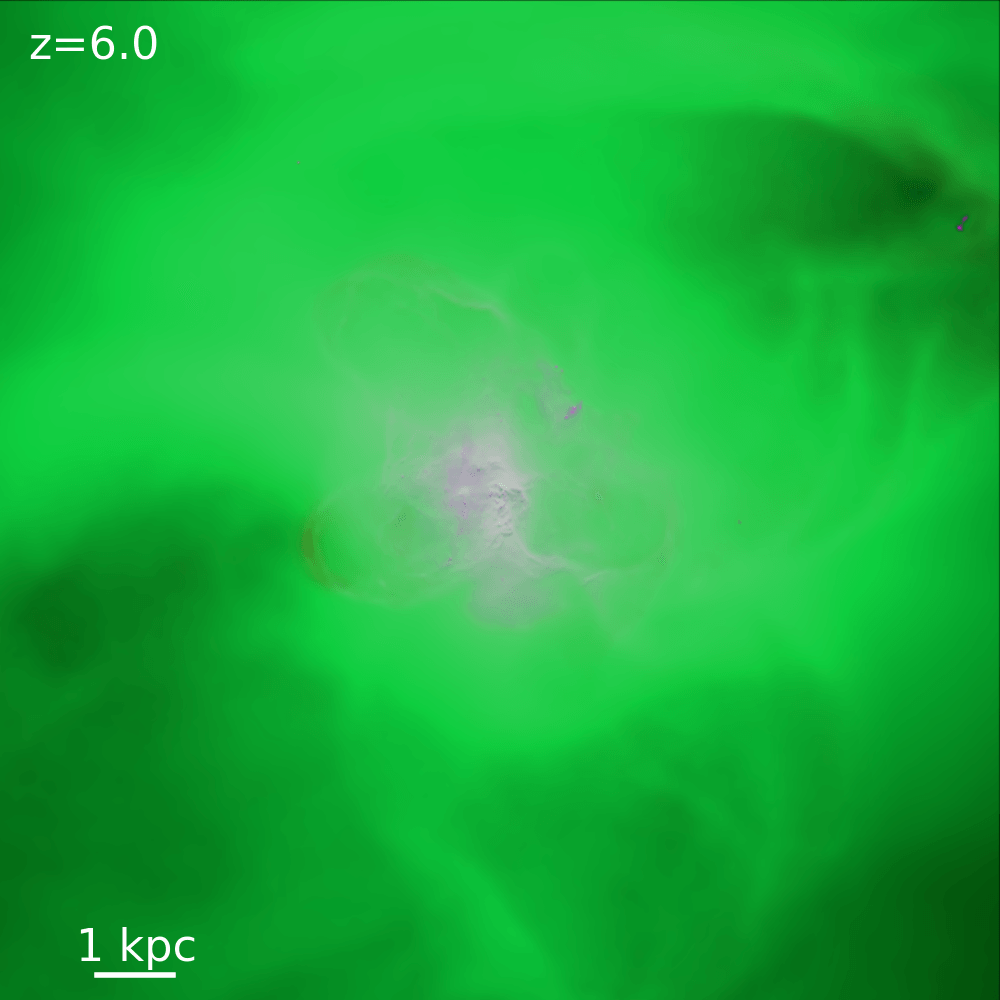} &
  \includegraphics[width=0.3\textwidth]{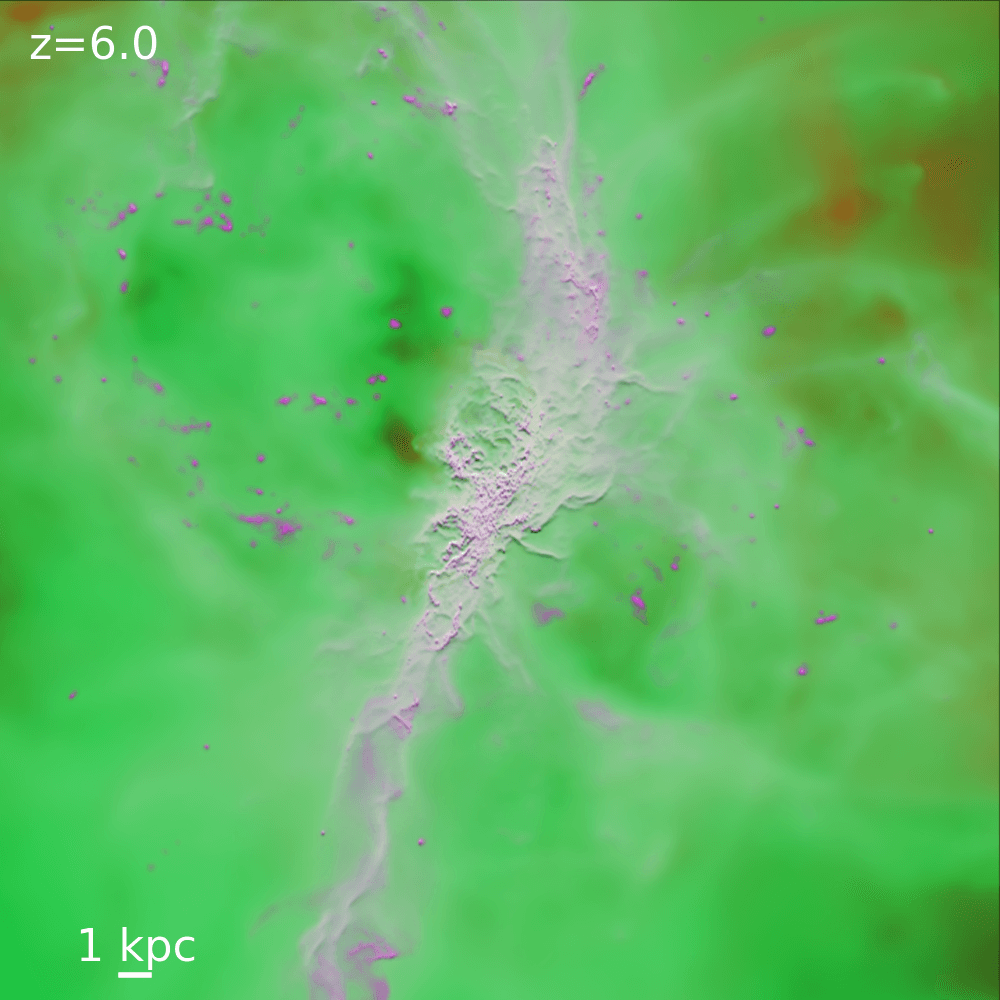} &
  \includegraphics[width=0.3\textwidth]{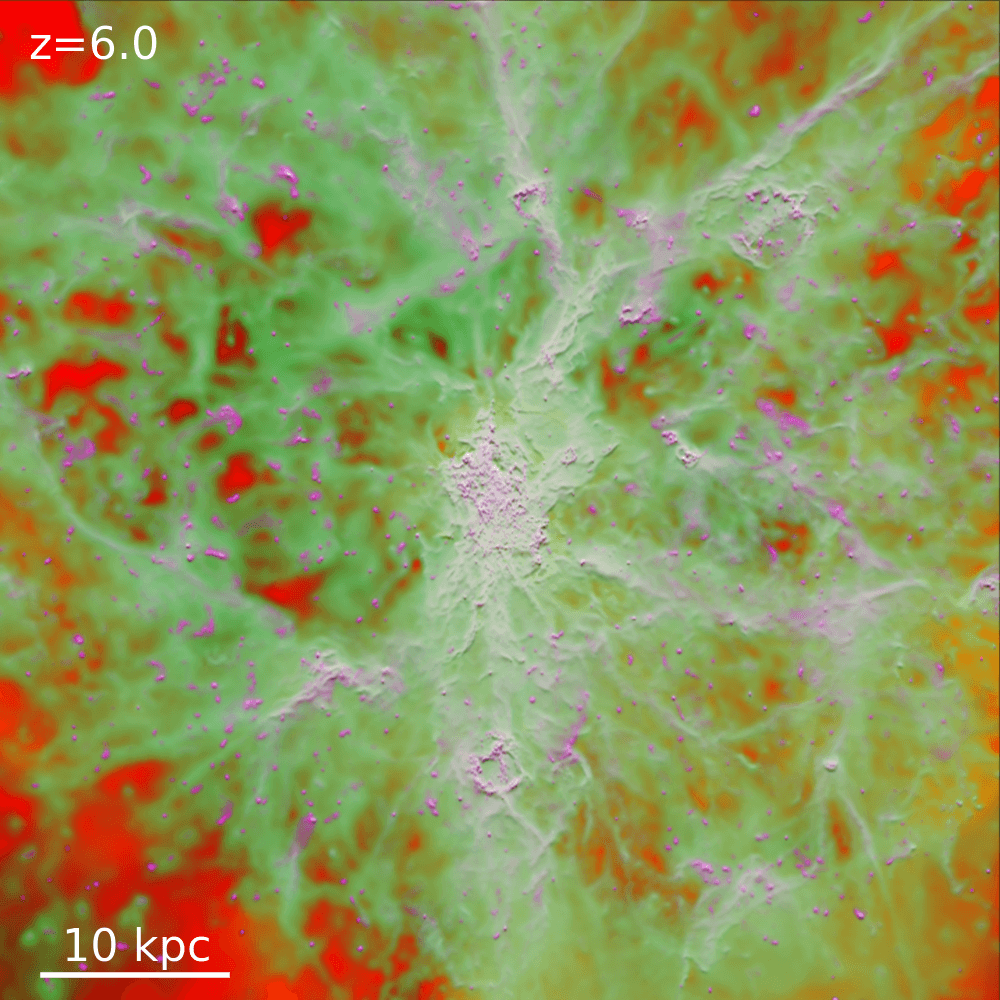} \\
  \includegraphics[width=0.3\textwidth]{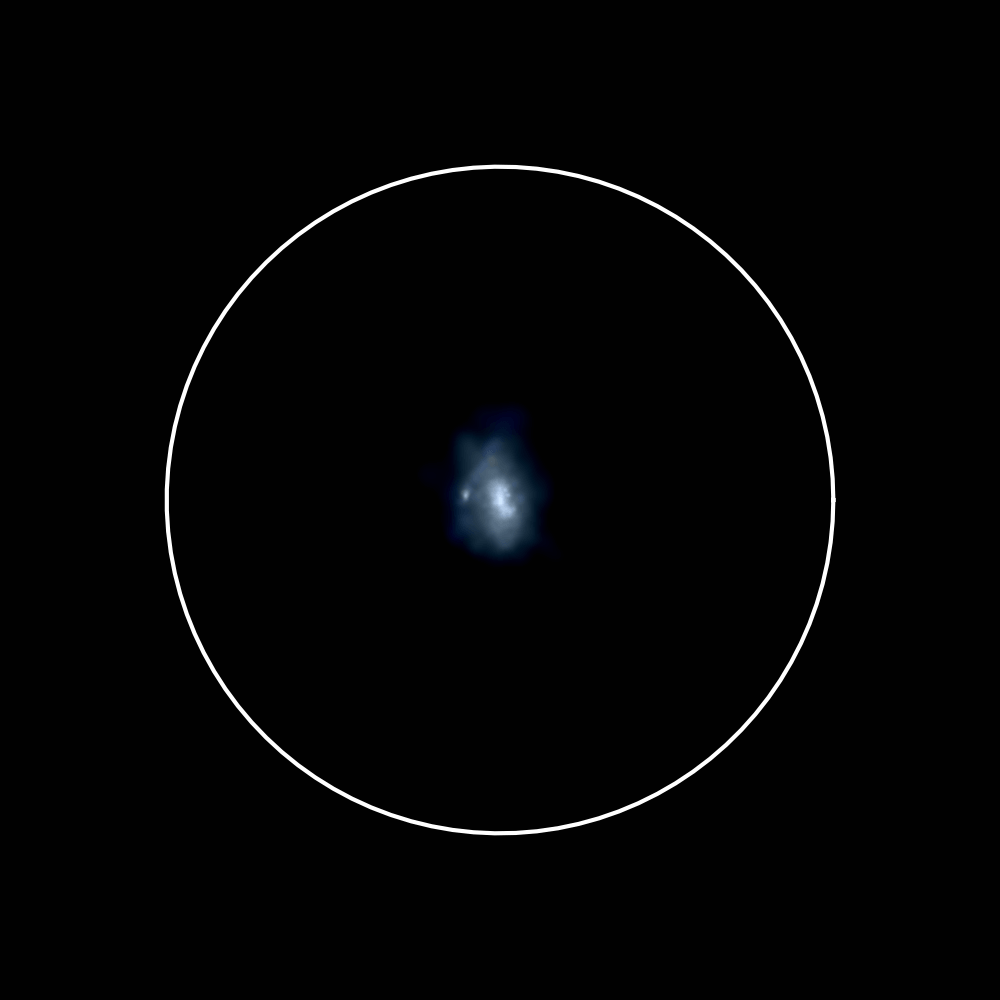} &
  \includegraphics[width=0.3\textwidth]{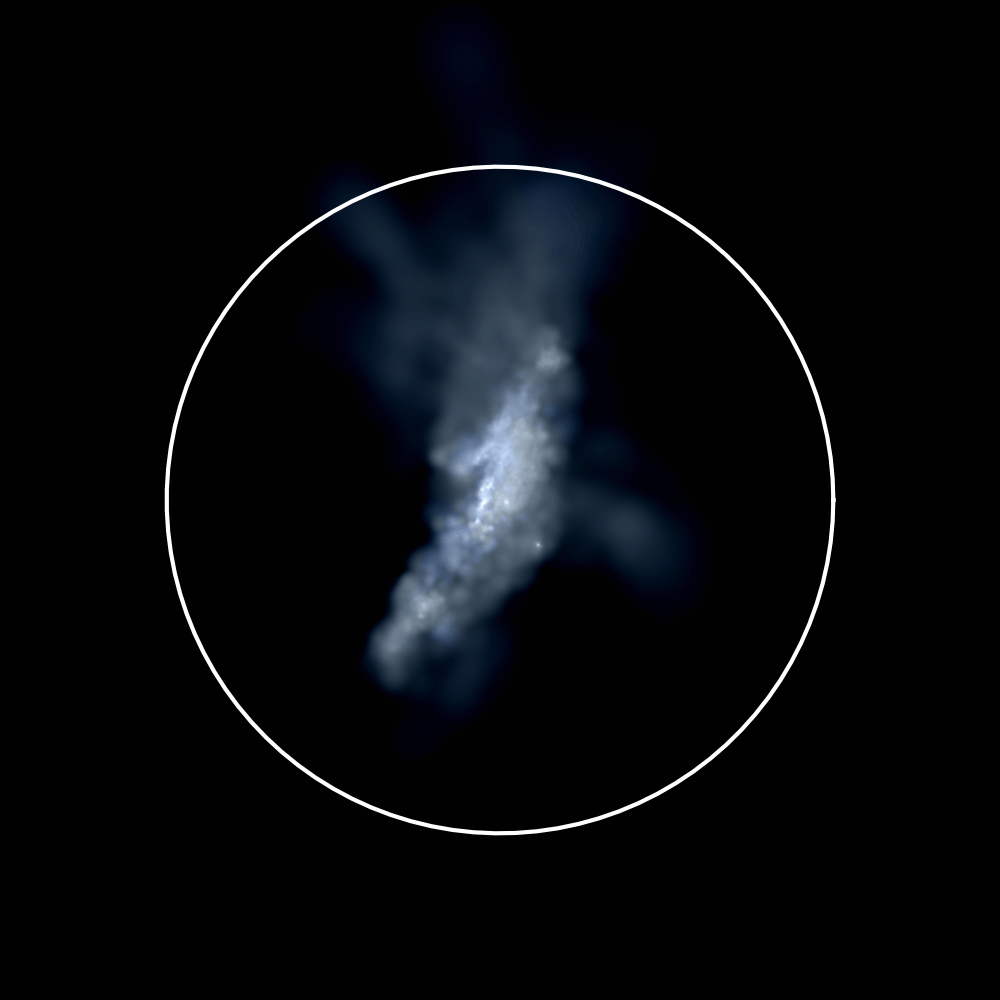} &
  \includegraphics[width=0.3\textwidth]{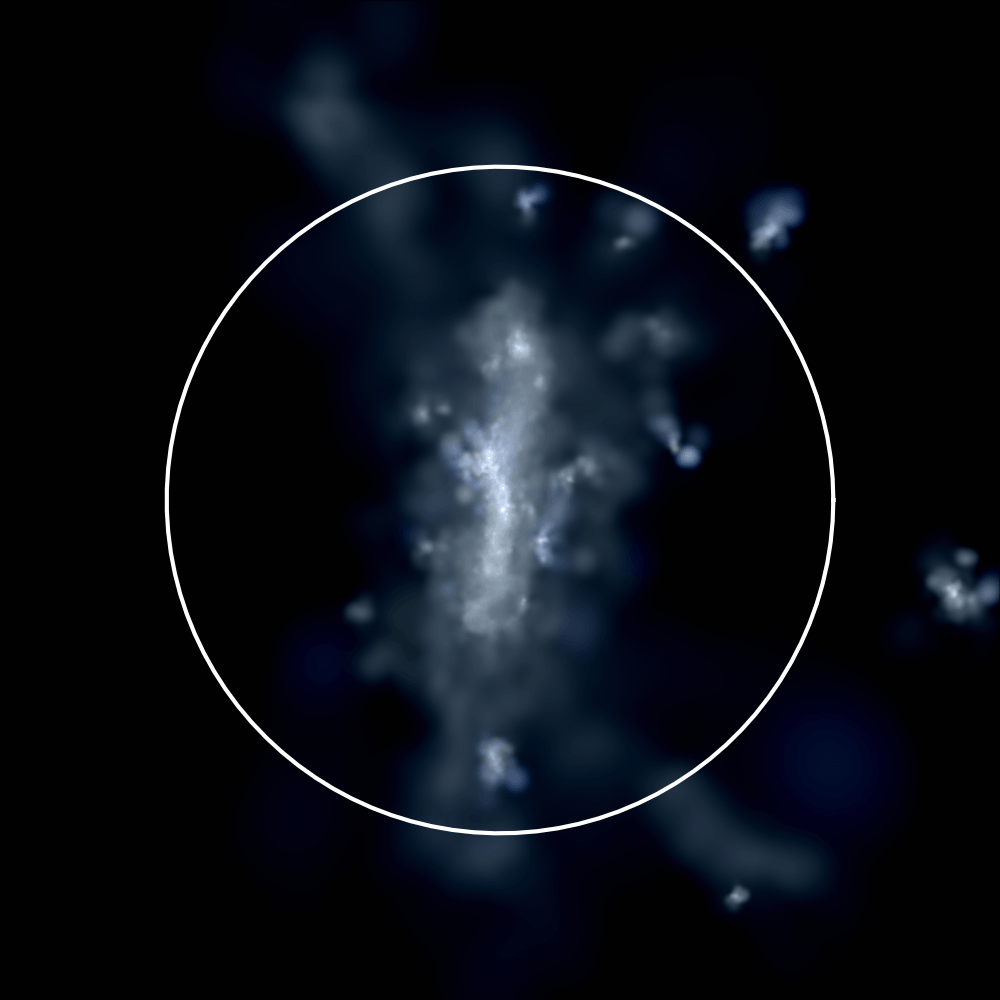} \\
\end{tabular}
\caption{Gas and stars in {\bf z5m09} (left column), {\bf z5m10mr} (middle column), and {\bf z5m11} (right column), at $z=9.6$ (upper panels) and $z=6.0$ (lower panels), respectively. Gas images show log-weighted projected gas density. Magenta shows cold molecular/atomic gas ($T<1000$ K), green shows warm ionized gas ($10^4\leq T\leq10^5$ K), and red shows hot gas ($T>10^6$ K) (see \citealt{hopkins.14.fire} for details). Stellar images are mock $u/g/r$ composites. We use STARBURST99 to determine the SED of each star particle from its known age and metallicity, and then ray-tracing the line-of-sight flux, attenuating with a MW-like reddening curve with constant dust-to-metals ratio for the abundance at each point. White circles show the position and halo virial radii of each main galaxy (see text) identified by the AHF code. Gas and star images of the same snapshot use the same projection and the same box size along each direction. We can clearly see a complicated, multi-phase ISM structure, with inflows, outflows, mergers, and star formation in dense clouds all occurring at the same time.}
\label{fig:image}
\end{figure*}

% Figure 2
\begin{figure*}
\centering
\begin{tabular}{ccc}
\centering
  Young & Middle-aged & Old \\
  \includegraphics[width=0.28\textwidth]{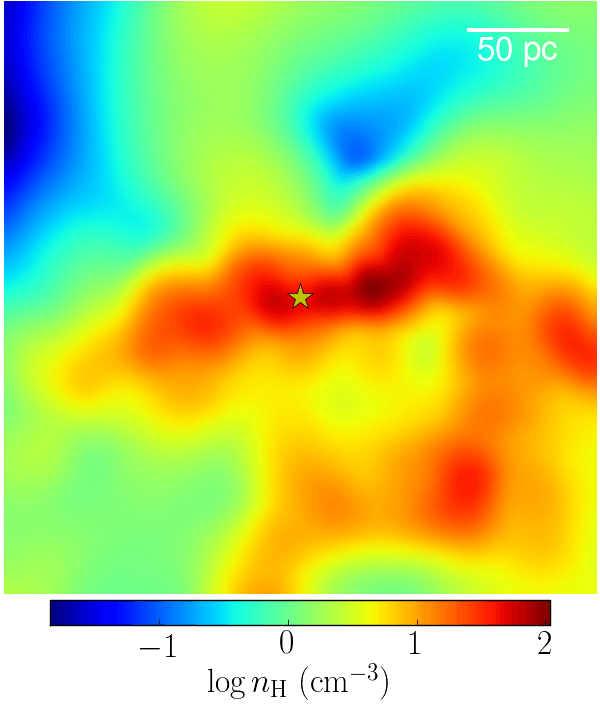} & 
  \includegraphics[width=0.28\textwidth]{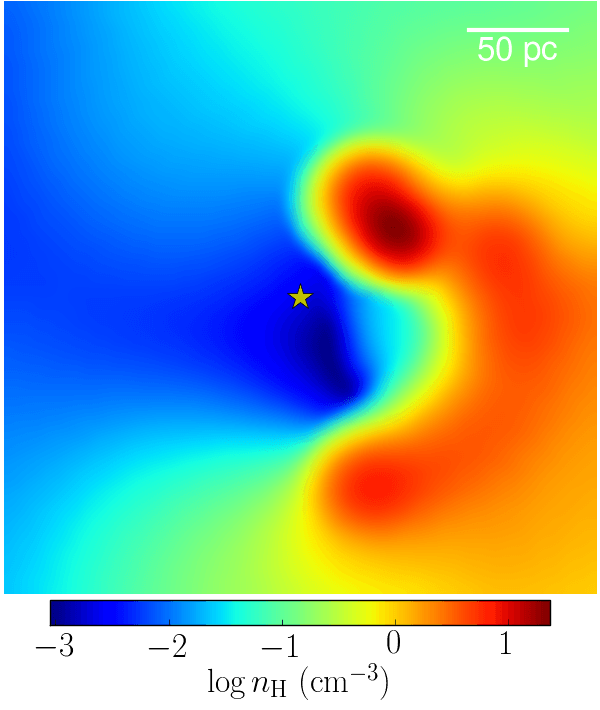} &
  \includegraphics[width=0.28\textwidth]{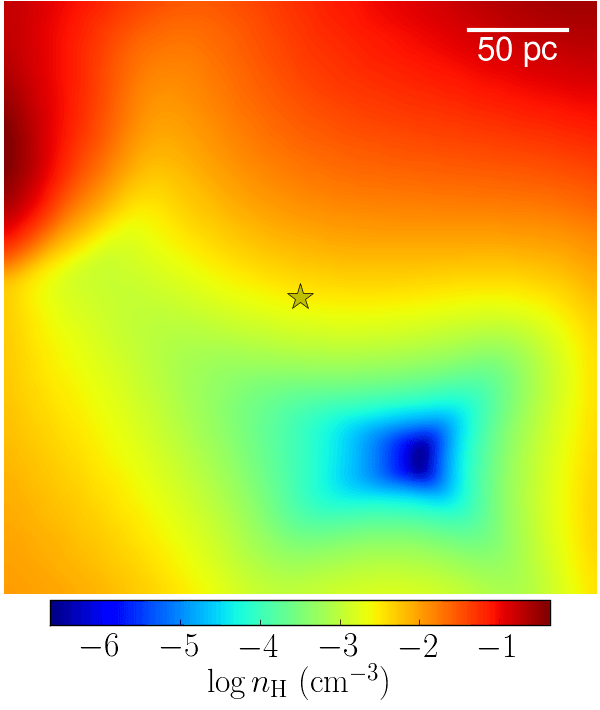} \\
  \includegraphics[width=0.28\textwidth]{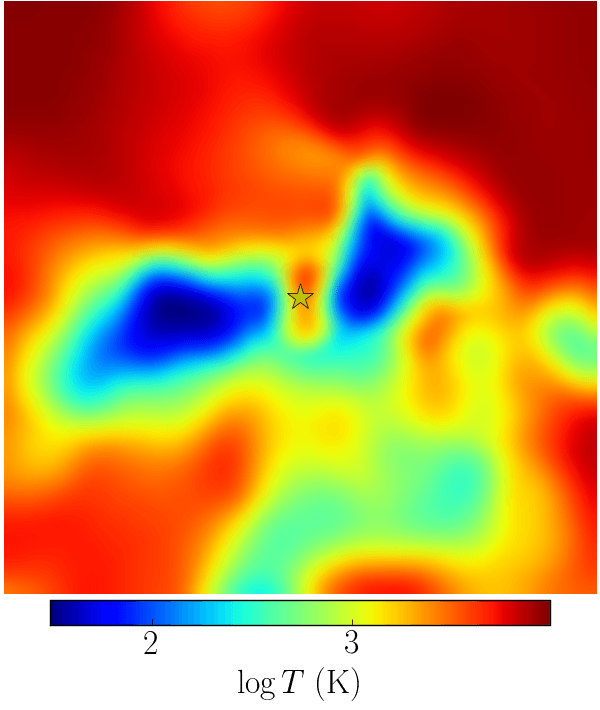} & 
  \includegraphics[width=0.28\textwidth]{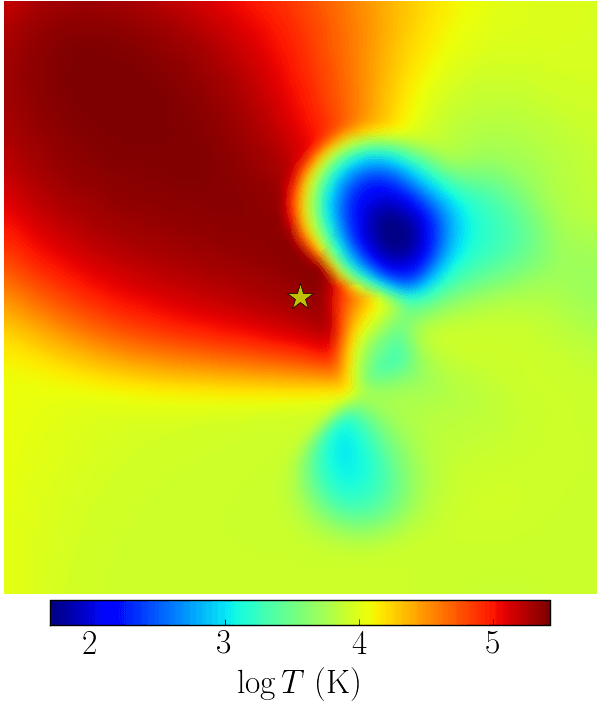} &
  \includegraphics[width=0.28\textwidth]{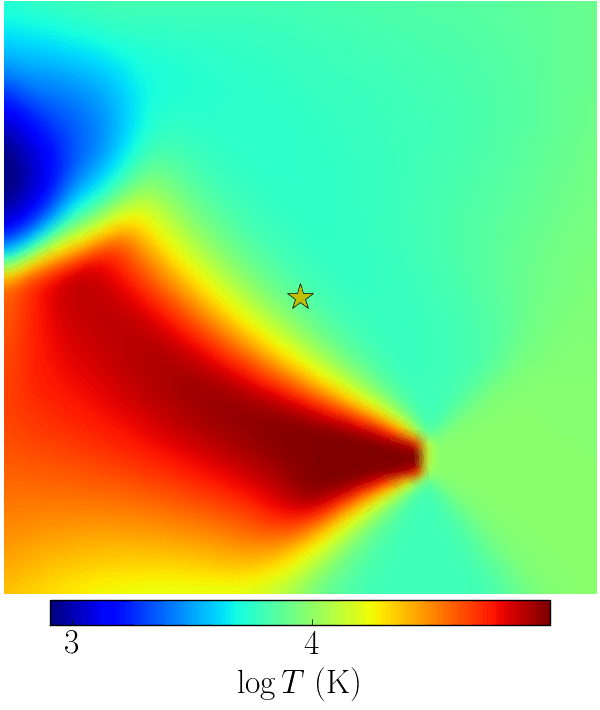} \\
\end{tabular}
\caption{ISM structure in a random star neighborhood. We show density (top panels) and temperature (bottom panels) maps of a slice around a(n) young ($\sim1$ Myr, left column), middle-aged ($\sim5$ Myr, middle column), and old ($\sim40$ Myr, right column) star particle. Each box is 300 pc along each direction. The yellow stars represent the position of the star particle. We clearly see that young stars -- which produce most of the ionizing photons -- are buried in H II regions inside their dense birth clouds. By $\gtrsim10$ Myr, the clouds are totally destroyed and most sightlines to the stars have low column densities, but these stars no longer produce many ionizing photons.}
\label{fig:ism}
\end{figure*}

\section{The Simulations}
\label{sec:simulations}
This work is part of the FIRE project \citep{hopkins.14.fire}. All the simulations use the newly developed GIZMO code \citep{hopkins.14.gizmo} in ``P-SPH'' mode. P-SPH adopts a Lagrangian pressure-entropy formulation of the smoothed particle hydrodynamics (SPH) equations \citep{hopkins.13.sph}, which eliminates the major differences between SPH, moving-mesh, and grid codes, and resolves many well-known issues in traditional density-based SPH formulations. The gravity solver is a heavily modified version of the GADGET-3 code \citep{springel.05.gadget}; and P-SPH also includes substantial improvements in the artificial viscosity, entropy diffusion, adaptive time-stepping, smoothing kernel, and gravitational softening algorithm. We refer to \citet{hopkins.13.sph,hopkins.14.gizmo} for more details on the numerical recipes and extensive test problems. A list of the simulations in this work is presented in Table~\ref{table:simlist}, while the parameters there will be introduced in the rest of this section.

The simulations in this work are of a separate series from other FIRE simulations. A large cosmological box was first simulated at low resolution down to $z=5$, and then halos of masses in $10^9$--$10^{11}\Msun$ at that time were picked and re-simulated in a smaller box at much higher resolution with the multi-scale ``zoom-in'' initial conditions generated with the MUSIC code \citep{hahn.11.music}, using second-order Lagrangian perturbation theory. The resolution of the simulations in this work can be roughly divided into two categories, which we refer as ``high resolution'' (HR) and ``medium resolution'' (MR), although the specific initial particle mass may vary according to the size of the system. Some initial conditions we adopt in the simulations and general properties of the galaxies at $z=6$ are listed in Table \ref{table:simlist}. We will show in Section \ref{sec:galaxy} that they are typical in most of their properties and thus can be considered as ``representative'' in this mass range.

In our simulations, gas follows an ionized+atomic+molecular cooling curve from 10--$10^{10}$ K, including metallicity-dependent fine-structure and molecular cooling at low temperatures and high-temperature metal-line cooling followed species-by-species for 11 separately tracked species \citep{wiersma.09.cooling}. We do not include a primordial chemistry network nor try to model the formation of Pop III stars, but apply a metallicity floor of $Z=10^{-4}Z_{\odot}$ in the simulations. Therefore, we will focus our analysis at $z\lesssim11$, when our galaxies are sufficiently metal-enriched.

At each timestep, the ionization states are determined from the photoionization equilibrium equations described in \citet{katz.96.ionization} and the cooling rates are calculated from a compilation of CLOUDY runs, by applying a uniform but redshift-dependent photo-ionizing background tabulated in \citet{fg.09.uvb}\footnote{The photo-ionizing background stars to kick in at $z=10.6$ and is available at http://galaxies.northwestern.edu/uvb/.}, and photo-ionizing and photo-electric heating from local sources. Gas self-shielding is accounted for with a local Jeans-length approximation, which is consistent with the radiative transfer calculation in \citet{fg.10.lya}. In this work, we also post-process the simulations with full radiative transfer calculation and re-compute the ionization states. We find consistent results between the post-processing and on-the-fly calculations (see Appendix \ref{sec:appendix} for details).

The models of star formation (SF) and stellar feedback implemented in the FIRE simulations are developed and presented in a series of papers \citep[][and references therein]{hopkins.11.fb,hopkins.12.fb,hopkins.13.sf,hopkins.14.fire}. We briefly summarize their main features here and refer to the references for more details and discussion. We follow the SF criteria developed in \citet{hopkins.13.sf} and allow stars to form only in molecular and self-gravitating gas clouds with number density above some threshold $\nc$. We choose $\nc=100~\cm^{-3}$ as the fiducial value. It corresponds to the typical density of GMCs and is much larger than the mean ISM density in our simulations\footnote{On the other hand, the threshold is much less than the highest density these simulations can resolve, to save computational expense.}. In {\bf z5m10h}, we set $\nc=1000~\cm^{-3}$ for a convergence test. SF occurs at 100\% efficiency per free-fall time when the gas meets these criteria (i.e. $\dot{\rho}_{\ast}=\rho/t_{\rm ff}$). This SF prescription adaptively selects the largest over-densities and automatically predicts clustered SF \citep{hopkins.13.sf}. It is also motivated by much higher-resolution, direct simulations of dense, star-forming clouds \citep{padoan.11.sf,evs.11.sf,federrath.11.sf}. A star particle inherits the metallicity of each tracked species from its parent gas particle, and its age is determined by its formation time in subsequent timesteps.

The {\bf z5m10e} run is intentionally designed to mimic ``sub-grid'' star formation models as commonly adopted in low-resolution simulations that cannot capture the star formation in dense gas clouds. In this run, we lower $\nc$ to $1~\cm^{-3}$ and allow stars to form at 2\% efficiency per free-fall time in {\it all} gas above $1~\cm^{-3}$ but not self-gravitating (still 100\% efficiency in self-gravitating gas). This will result in a wide spatial and density distribution of SF and means that stars do not need to form in high-density structures. 

Every star particle is treated as a single stellar population with known age, metallicity, and mass. Then all the quantities associated with feedback, including ionizing photon budgets, luminosities, stellar spectra, supernovae (SNe) rates, mechanical luminosities of stellar winds, metal yields, etc., are directly tabulated from the stellar population models in STARBURST99 \citep{leitherer.99.sb}, assuming a \citet{kroupa.02.imf} initial mass function (IMF) from 0.1--$100~\Msun$. In principle, this ``IMF-averaged'' approximation breaks down in our HR simulations, where the mass of a star particle is only 10--$100~\Msun$. Previous studies showed that it has little effect on global galaxy properties \citep[e.g.][and references therein]{hopkins.14.fire}. We also test and confirm that this approximation does not affect our results on escape fraction (see Section \ref{sec:escfrac}). 

We account for different mechanisms of stellar feedback, including: {\bf (1)} local and long-range momentum flux from radiative pressure; {\bf (2)} energy, momentum, mass and metal injection from SNe and stellar winds; and {\bf (3)} photoionization and photoelectric heating. We apply the Type-II SNe rates from STARBURST99 and Type-Ia SNe rates following \citet{mannucci.06.sne}, when a star particle is older than 3 Myr and 40 Myr, respectively. We assume that every SN ejecta has an initial kinetic energy of $10^{51}$ ergs, which is coupled to the gas as either thermal energy or momentum, depending on whether the cooling radius can be resolved \citep[see][for more details]{hopkins.14.fire,martizzi.15.sne}. We also follow \citet{wiersma.09.metal} and adopt Type-II SNe yields from \citet{woosley.95.metal} and Type-Ia yields from \citet{iwamoto.99.snia}. We do not model SNe and metal enrichment from Pop III stars.

We emphasize that the on-the-fly photoionization is treated in an approximate way in our simulations -- we move radially outwards from the star and ionize each nearest neutral gas particle until the photon budget is exhausted. This treatment allows ionizing regions to overlap and expand, and is qualitatively reasonable in intense star-forming regions. However, when the gas distribution is highly asymmetric around an isolated star particle, their ionization states might not be accurately captured in the simulations. Nonetheless, as we will post-process our simulations with full radiative transfer code to trace the propagation of ionizing photons and re-compute the ionization states (Section \ref{sec:escfrac}), this approximation will have little effect on the escape fraction we evaluate. Also, in the region where the gas density is extremely high, photoionization may not be well-captured due to resolution limits. But we confirm that this neither has strong dynamical effect on gas structure in high-density regions nor affects the escape fraction\footnote{In our simulations, star particles have similar mass to gas particles. Applying a \citep{kroupa.02.imf} IMF and standard stellar population model, in regions with $\nc\sim100~\cm^{-3}$, the ionizing photons emitted from a young star particle can ionize the mass of two gas particles. However, some clouds reach densities $\gtrsim2000~\cm^{-3}$, where one needs to collect the ionizing photon budgets from 10 young star particles to fully ionize a single gas particle. In the code, the on-the-fly estimate of HII photoionization feedback treats this limit stochastically \citep[see][]{hopkins.11.fb}, so we might risk underestimating the dynamical effects of photo-heating. Therefore, we run a simulation where we artificially boost the ionizing photon budget by a factor of $10$, which is not physical but dramatically reduces the stochastic variations. We find that the typical gas density of star-forming clouds and the average escape fractions (computed from our post-processing radiative transfer, see Section \ref{sec:escfrac}) are very similar to our standard runs. Therefore, we confirm that the on-the-fly photoionization feedback approximation in our simulations does not strongly affect our results.}.

The simulations described in Table \ref{table:simlist} adopt a standard flat $\Lambda$CDM cosmology with cosmological parameters $H_0=70.2 {\rm~km~s^{-1}~Mpc^{-1}}$, $\Omega_{\Lambda}=0.728$, $\Omega_{m}=1-\Omega_{\Lambda}=0.272$, $\Omega_b=0.0455$, $\sigma_8=0.807$ and $n=0.961$, which are within the uncertainty of current observations \citep[e.g.][]{hinshaw.13.wmap,planck.13}.

\section{Galaxy Properties}
\label{sec:galaxy}
\subsection{Halo Identification}
The galaxies in our simulations have different assembly histories at high redshifts. The smallest galaxy, {\bf z5m09}, evolves primarily via accretion and passive evolution, while the more massive ones have undergone multiple mergers at earlier times. We use the Amiga Halo Finder \citep[AHF;][]{gill.04.ahf,knollmann.09.ahf} to identify halos in the simulations. The AHF code uses an adaptive mesh refinement method. We choose the centre of a halo as the centre of mass of all particles in the finest refinement level and adopt the virial overdensity from \citet{bryan.norman.98}. In this work, we only consider the main galaxies that are well-resolved in the simulations. We exclude those that are contaminated by low-resolution paticles, not sufficiently resolved (contain less than $10^5$/$10^6$ bound particles in MR/HR runs, or have stellar mass lower than $10\%$ of the most massive galaxy in each snapshot), and subhalos/satellite galaxies. Some example images of gas and stars at different redshifts are presented in Figure~\ref{fig:image}. The white circles show the virial radius of each halo. As the figure shows, the more massive systems were assembled by merging several smaller halos at early time.

\subsection{Multi-phase ISM Structure}
One advantage of our simulation is that we are able to explicitly resolve a realistic multi-phase ISM, star formation, and stellar feedback. Figure \ref{fig:image} shows the distribution cold, warm, and hot phase of gas on galactic scale. In Figure \ref{fig:ism}, we show some examples of ISM structure on sub-kpc scale around star particles of different ages from {\bf z5m10mr}. The left column is the density and temperature maps around a star particle of age 1 Myr (before the first SNe explode at $\sim3$ Myr). As expected from our star formation criteria, newly formed stars are embedded in their dense ``birth'' clouds. Within the central few pc around the star particle, the dense gas is ionized and heated by ionizing photons from the star and an H II region forms\footnote{For a typical gas density of 100 $\cm^{-3}$ and an ionizing photo budget 10$^{49.5}$ s$^{-1}$ in this simulation, the Str{\" o}mgren radius is around 5 pc.}. The middle column shows the ISM structure around an intermediate-age star particle (3--10 Myr), where there has just been a SN explosion (the example is 5 Myr old). The birth cloud has been largely dispersed and cleared by radiation pressure and SN feedback, opening a large covering fraction of low-density regions. In contrast, old star particles (right column, $\sim 40$ Myr) tend to be located in a warm, ambient medium. The ISM structures around star particles of different ages are very important in understanding the propagation of ionizing photons.

% Figure 3
\begin{figure}
\centering
\includegraphics[width=0.5\textwidth]{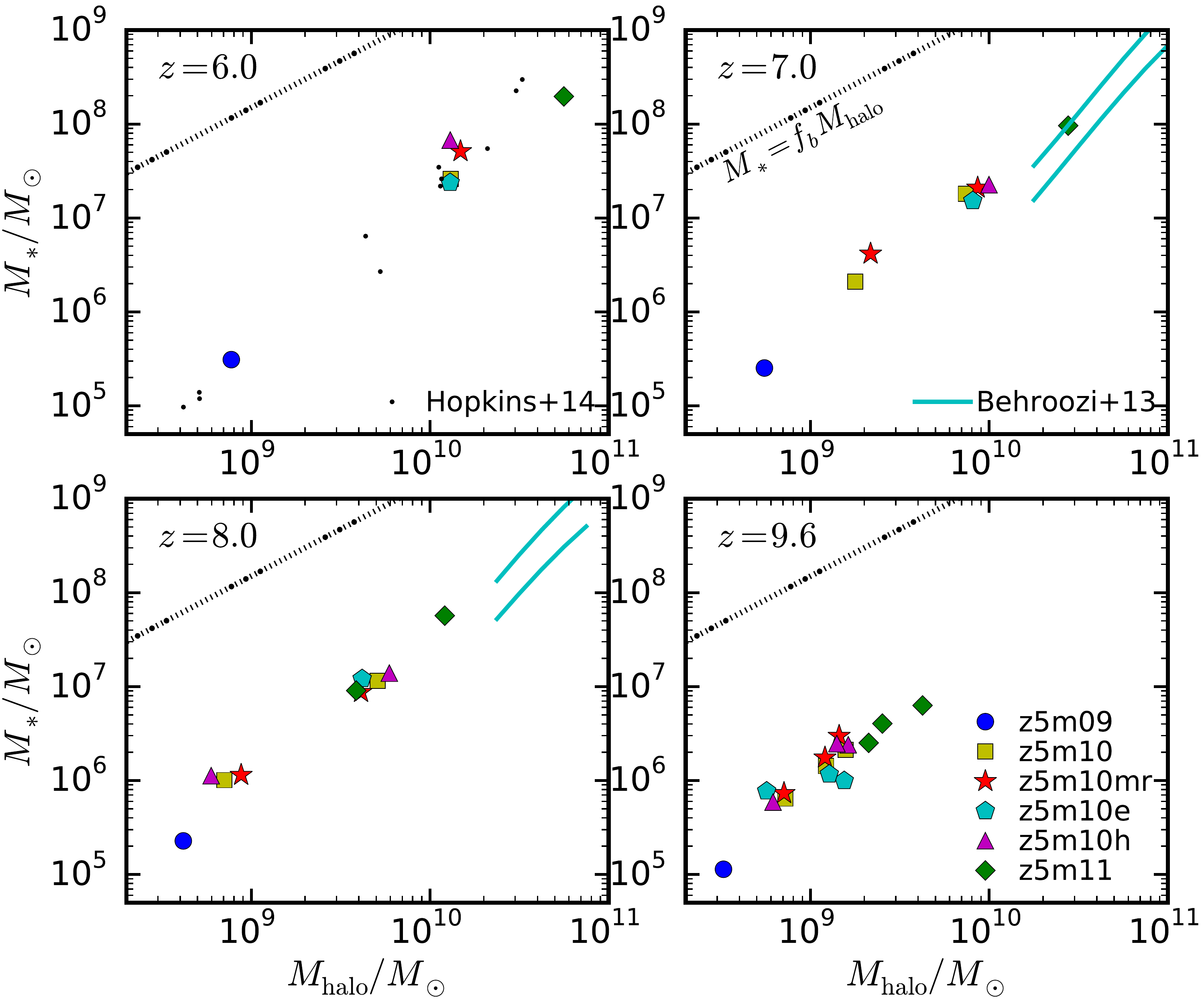}
\caption{Galaxy stellar mass--halo mass relation at $z=6$, 7, 8, and 9.6. We compare the relation with the simulations from \citet{hopkins.14.fire} at $z=6$ (small black dots), and the observationally inferred relation in \citet[][$z=7.0$ and $z=8.0$ only, cyan lines]{behroozi.13.sfh}. The black dotted lines represent the relation if all baryons turned into stars (i.e. $M_{\ast}=f_b~M_{\rm halo}$). Our simulations are broadly consistent with observations. These simulations are consistent with those in \citet{hopkins.14.fire}, although the latter have much lower resolution. It is reassuring that the stellar mass is converged in these runs.}
\label{fig:msmh}
\end{figure}

% Figure 4
\begin{figure}
\centering
\includegraphics[width=0.5\textwidth]{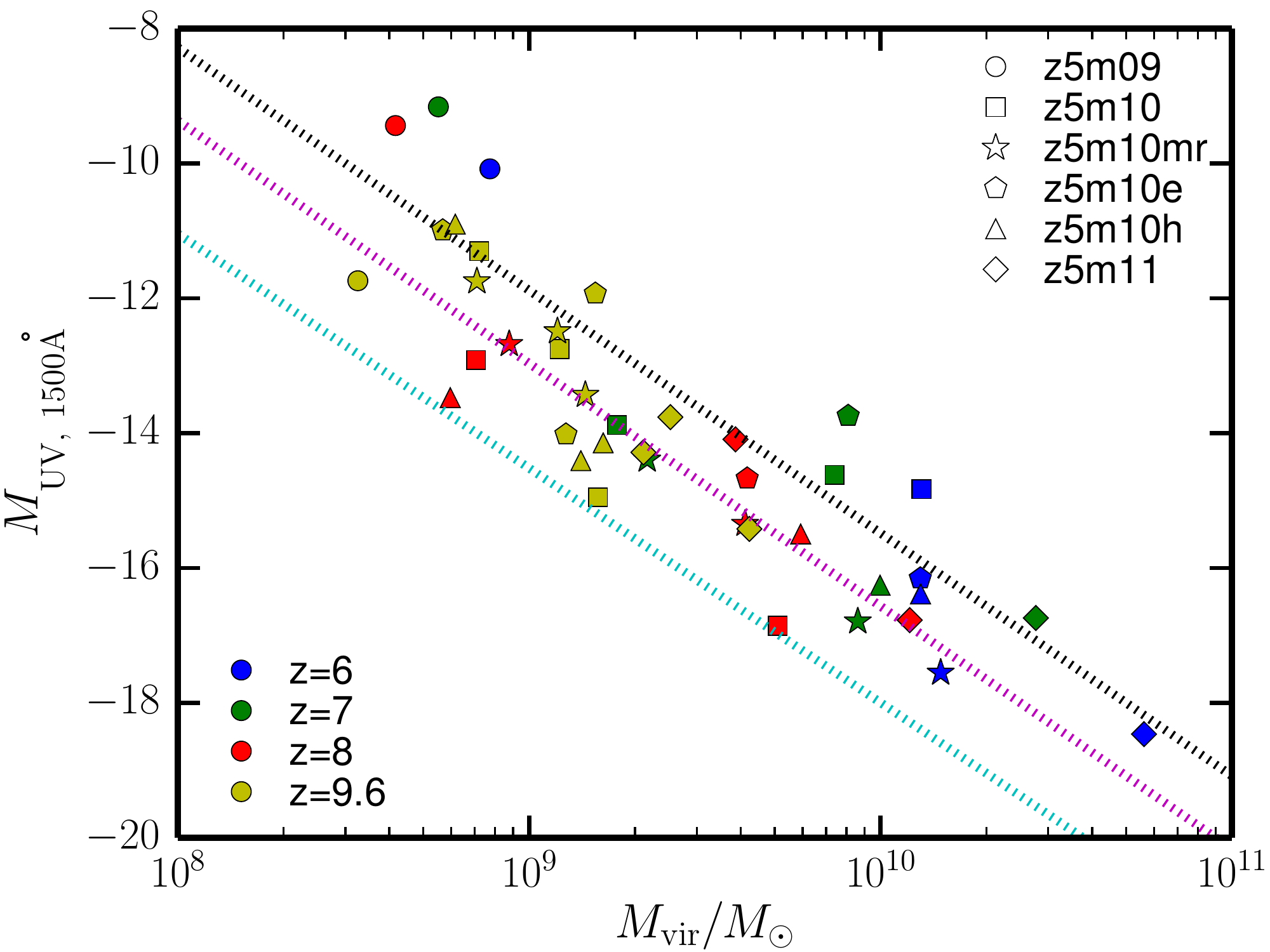}
\caption{UV magnitude at 1500 \AA~ as a function of halo mass for the simulated galaxies, color coded by redshift. Galaxies at $z=$6, 7, 8, and 9.6 are shown by blue, green, red, and yellow points, respectively. The numbers are calculated by converting the intrinsic luminosity at 1500 \AA~ to absolute AB magnitude. Dotted lines show the abundance matching from \citet[][fig 3]{kuhlen.fg.12} at $z=4$ (black), 7 (magenta), and 10 (cyan). The simulations are qualitatively consistent with the abundance matching, and span the range of $\muv=-9$ to -19 that is believed to dominate reionization.}
\label{fig:mhmag}
\end{figure}

% Figure 5
\begin{figure}
\centering
\includegraphics[width=0.5\textwidth]{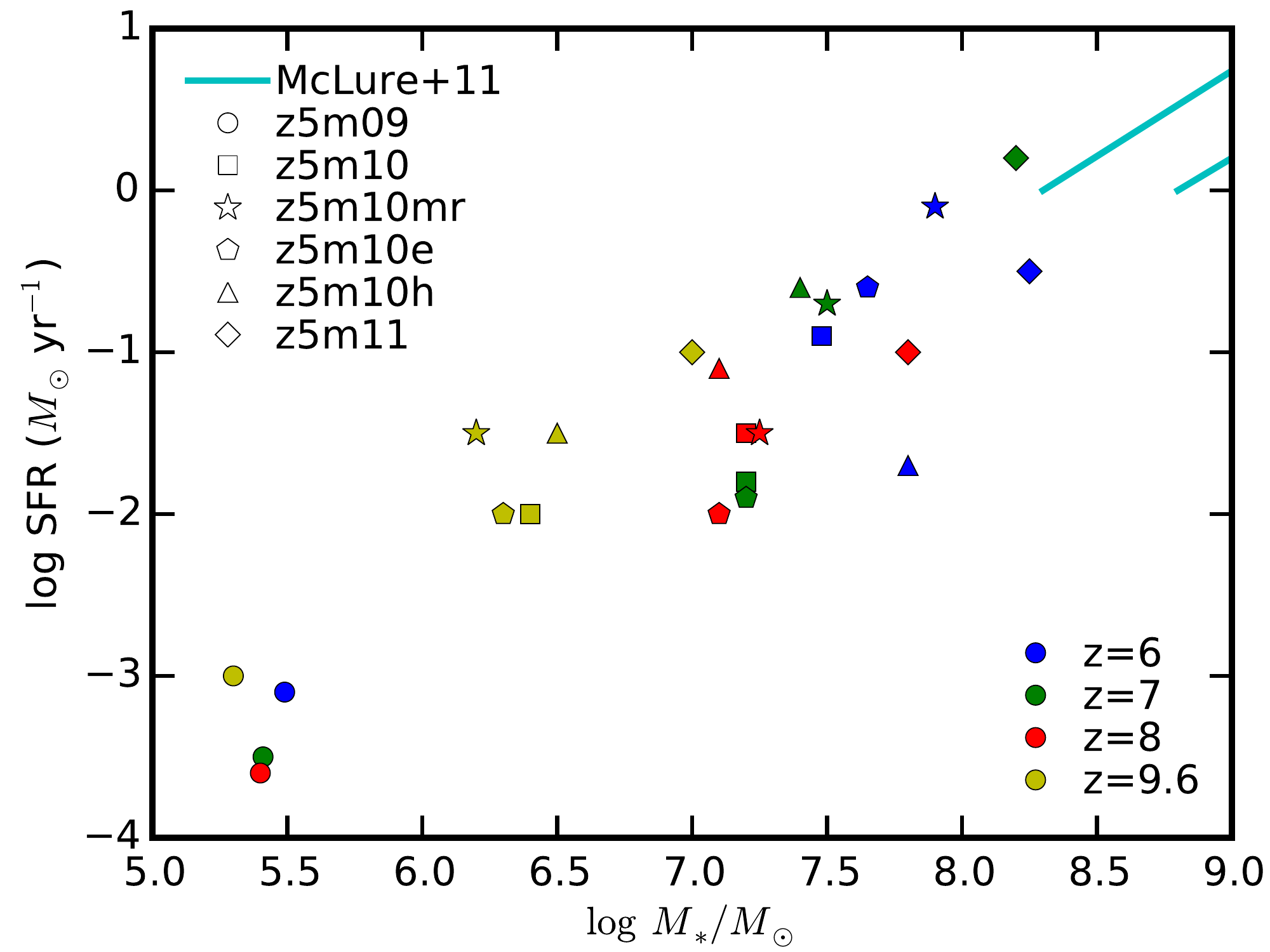}
\caption{Star formation rate--stellar mass relation at $z=6$, 7, 8, and 9.6 for the most massive galaxy in each simulation. The cyan lines illustrate the observed relation from a sample of galaxies at $z=6$--8.7 in \citet{mclure.11.sfr}. Our simulated galaxies agree with the observed relation where they connect at $\log(M_{\ast}/\Msun)=8.25$.}
\label{fig:sfr}
\end{figure}

\subsection{Galaxy Masses, Stellar Mass Assembly, and Star Formation History}
As has been shown in \citet{hopkins.14.fire}, with the stellar feedback models described here (with no tuned parameters), the FIRE simulations predict many observed galaxy properties from $z=0$--6: the stellar mass--halo mass relation, the Kennicutt--Schimidt law, star formation histories (SFHs), and the star-forming main sequence. The simulations in this work are of much higher resolution and focus on higher redshifts than those in \citet{hopkins.14.fire}. We extend their analysis and present the stellar mass-halo mass relation at $z=6$, 7, 8, and 9.6 for our simulated galaxies in Figure~\ref{fig:msmh}. We compare our results with the simulations from \citet{hopkins.14.fire} at $z=6$ and the observationally inferred relation from \citet{behroozi.13.sfh} at $z=7$ and 8 (note the relation in \citealt{behroozi.13.sfh} at $z=6$ does not overlap with the halo masses presented here). We confirm that our simulations predict stellar masses consistent with observations at these redshifts. It is also reassuring that the stellar masses in these simulations are well converged, despite those from \citet{hopkins.14.fire} having much poorer resolution.

In Figure \ref{fig:mhmag}, we present the relation between UV magnitude at rest-frame 1500 {\AA} and halo mass for our simulated galaxies at $z=6$, 7, 8, and 9.6. To obtain the UV magnitudes, we first calculate the specific luminosity at 1500 {\AA} for each star particle by interpolating the stellar spectra tabulated from STARBURST99 as a function of age and metallicity, and then convert the galaxy-integrated luminosity to absolute AB magnitude. In Figure \ref{fig:mhmag}, we also compare with the interpolated abundance matching from \citet[][dotted lines]{kuhlen.fg.12}. The simulations are qualitatively consistent with the abundance matching results, and lie within the systematic observational uncertainties. Given that in this simple calculation, we ignore the attenuation from dust inside the galaxy and along the line-of-sight in the IGM, and that the abundance matching is very uncertain at the faint end, we do not further discuss the comparison with these results. The simulated galaxies cover $\muv=-9$ to $-19$ at these redshifts, which are believed to play a dominant role in providing ionizing photons during reionization \citep[e.g.][]{finkelstein.12.candel,kuhlen.fg.12,robertson.13.udf12}. Most of these galaxies are too faint to be detectable in current observations; our {\bf z5m11} galaxy is, however, just above the detection limit ($\muv\sim-18$) of many deep galaxy surveys beyond $z\sim6$. Next generation space and ground-based facilities, such as the {\it James Webb Space Telescope} ({\it JWST}) and the {\it Thirty Meter Telescope} ({\it TMT}) may push the detection limit down to $\muv\sim-15.5$ \citep[e.g.][]{wise.14} and many of our simulated galaxies will then lie above the detection limits of future deep surveys.

\begin{figure*}
\centering
\begin{tabular}{ccc}
\centering
  \includegraphics[width=0.33\textwidth]{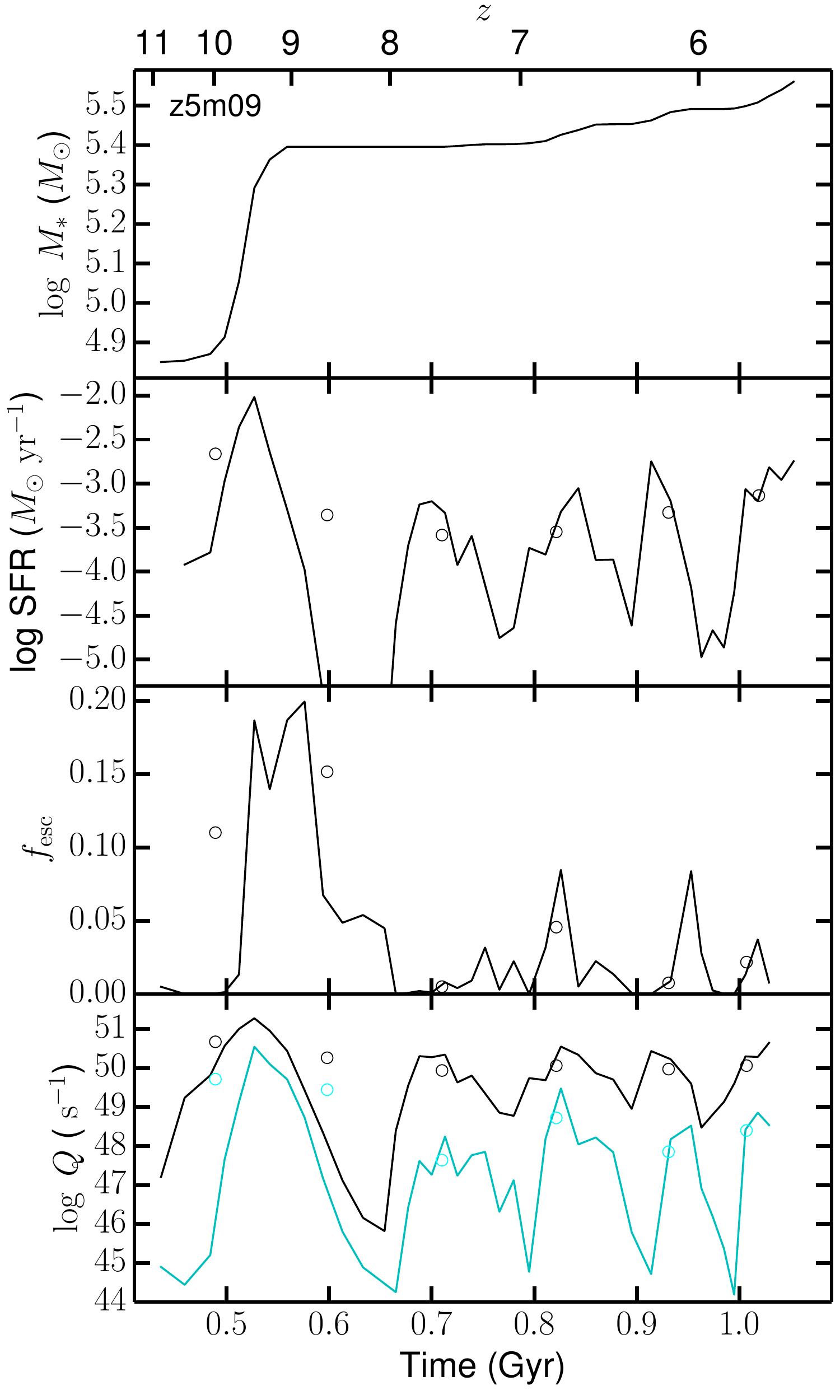} & 
  \includegraphics[width=0.33\textwidth]{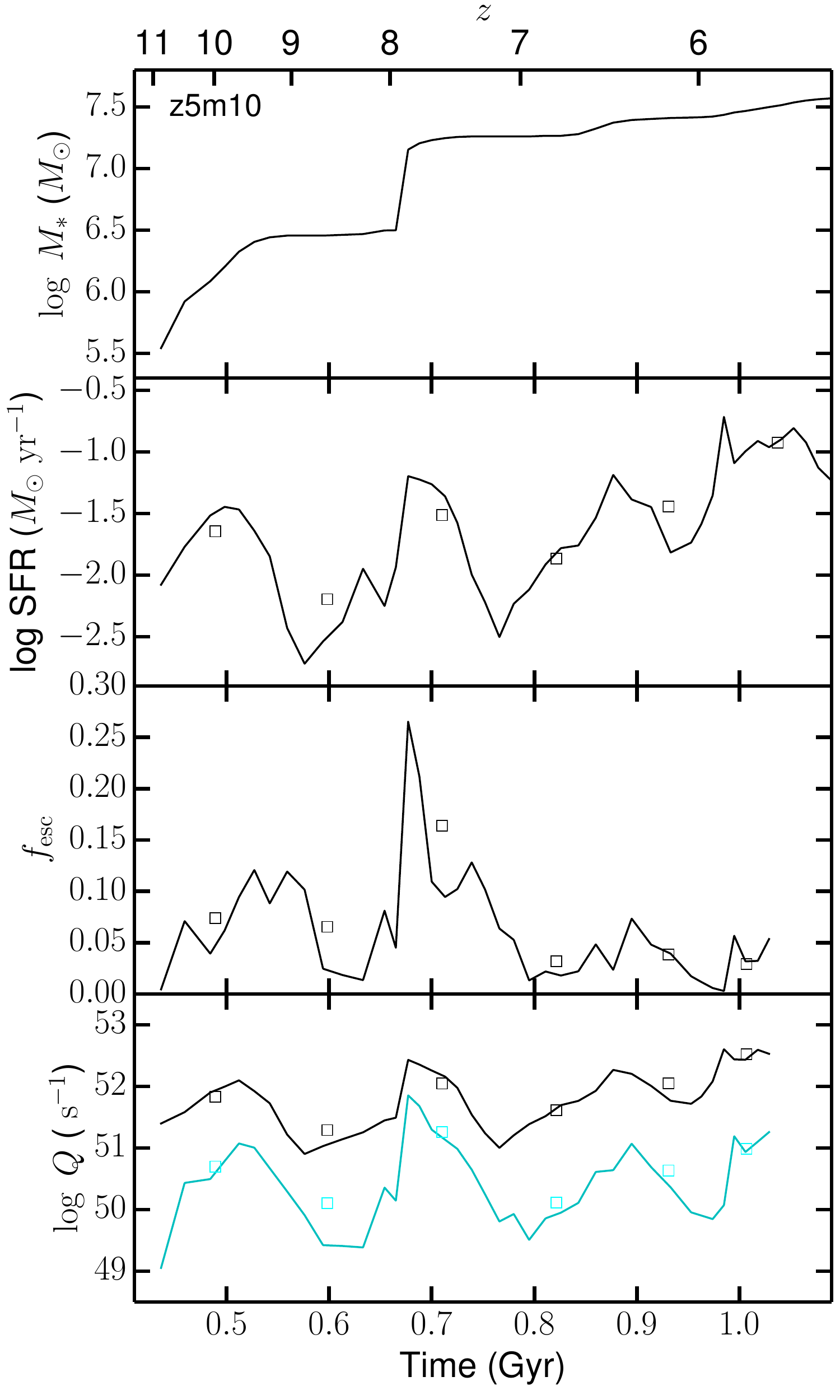} &
  \includegraphics[width=0.33\textwidth]{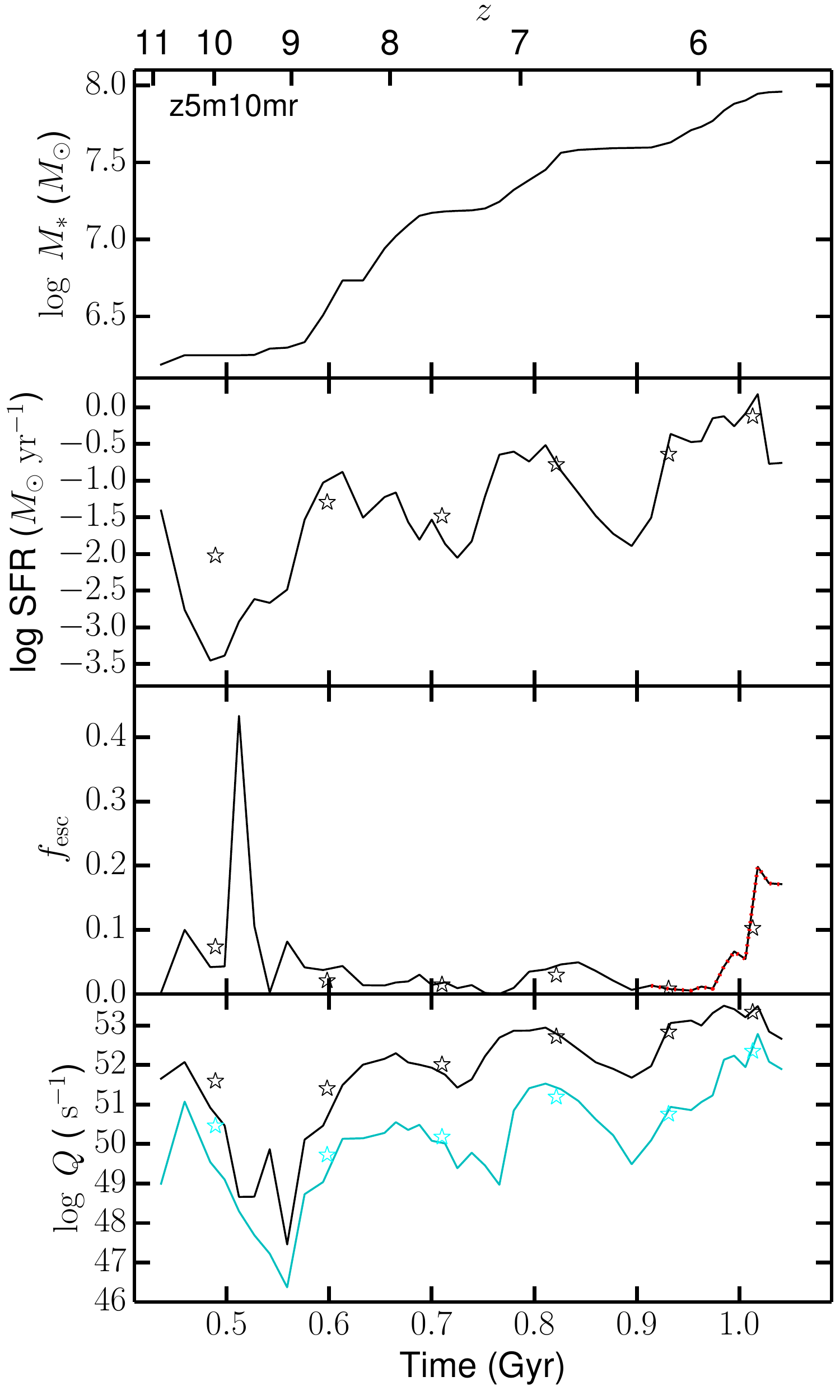} \\
  \includegraphics[width=0.33\textwidth]{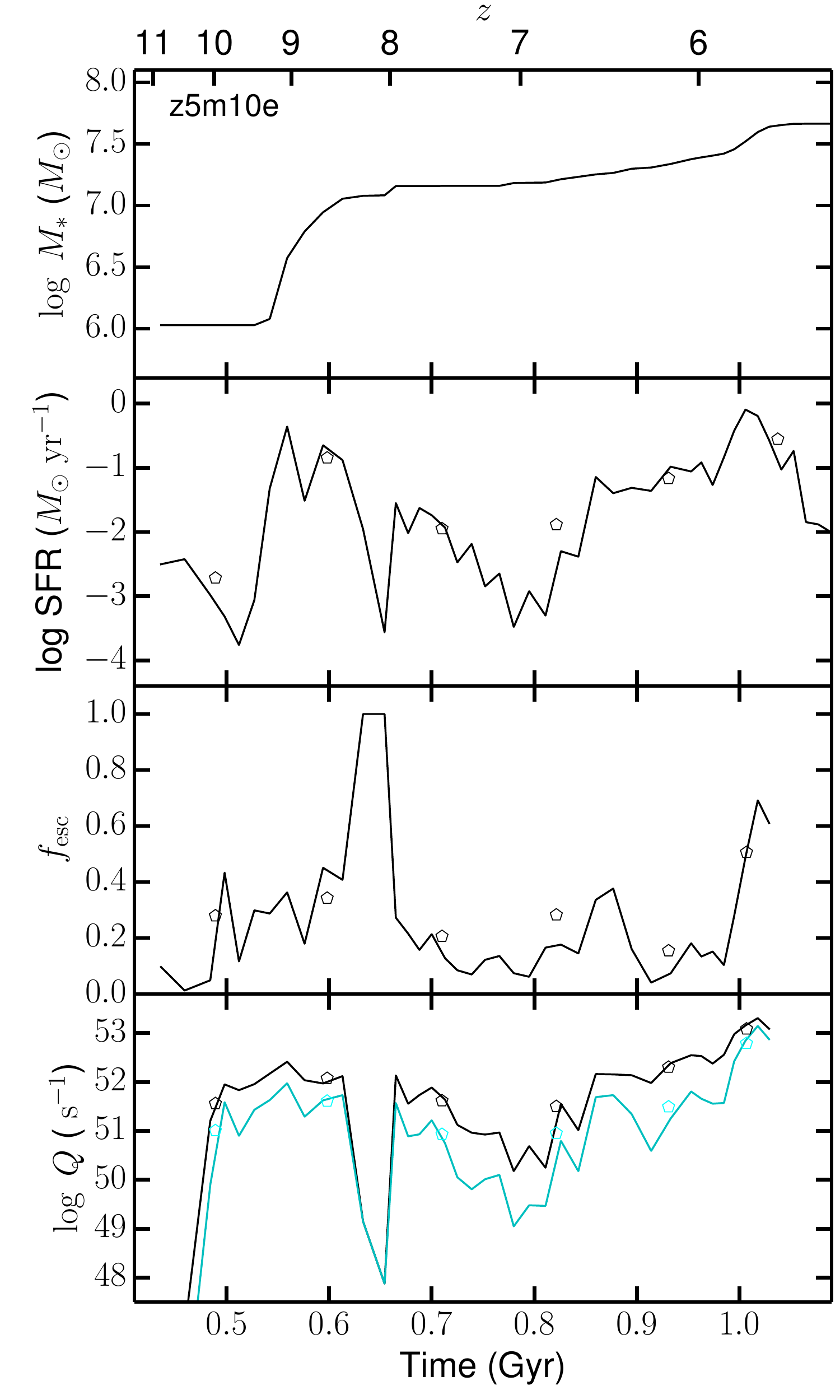} & 
  \includegraphics[width=0.33\textwidth]{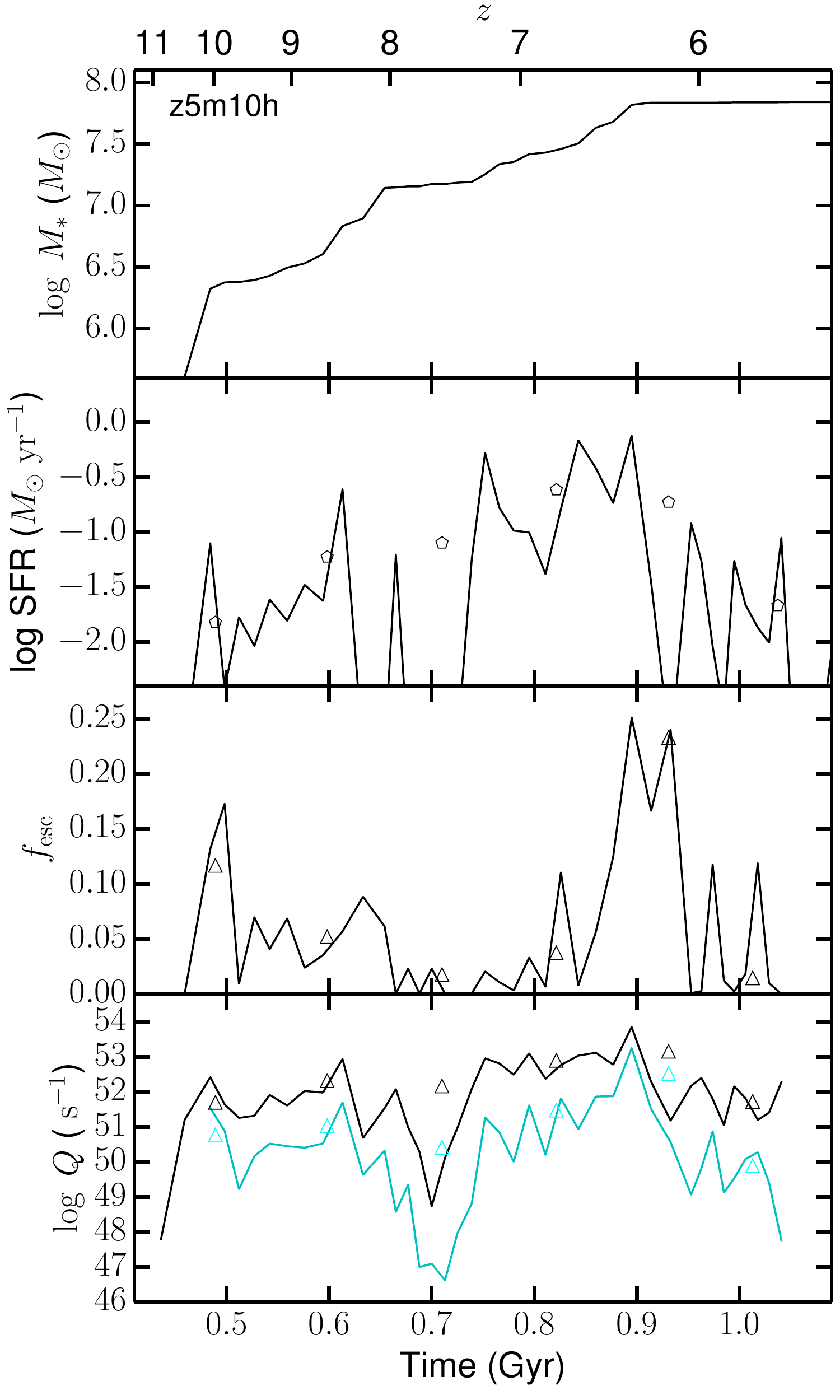} & 
  \includegraphics[width=0.33\textwidth]{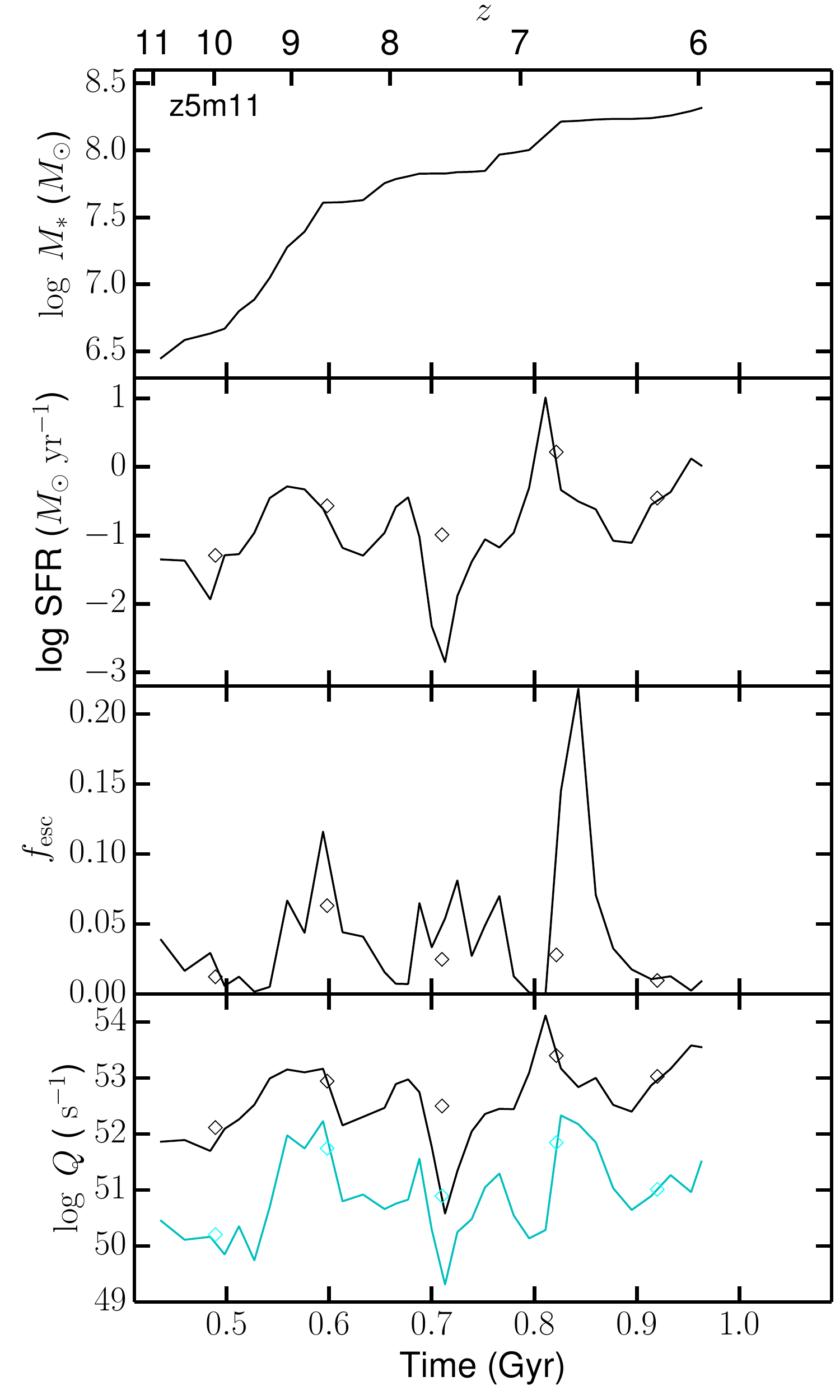} \\
\end{tabular}
\caption{Stellar mass (top panels), star formation rate (second panels), escape fraction (third panels), and intrinsic and escaped ionizing photon budget (bottom panels, black and cyan lines, respectively) as a function of cosmic time for the most massive galaxy in each run. Open symbols show the time-averaged quantities over 100 Myr. The red dotted line in {\bf z5m10mr} shows the escape fraction calculated with the UV background turned off (Section \ref{sec:uvb}). Instantaneous escape fractions are highly time variable, while the time-average escape fractions (over time-scales 100--1000 Myr) are modest ($\sim5\%$). The intrinsic ionizing photon budgets are dominated by stellar population younger than 3 Myr, which tend to be embedded in dense birth clouds. Most of the escaping ionizing photons come from stellar populations aged 3--10 Myr, where a large fraction of sightlines have been cleared by stellar feedback. Note that the run using a common ``sub-grid'' star formation model ({\bf z5m10e}), which allows stars to form in diffuse gas instead of in dense clouds, severely over-estimates $\fesc$.}
\label{fig:escfrac}
\end{figure*}

Figure \ref{fig:sfr} shows the star formation rate--stellar mass relation at $z=6$, 7, 8, and 9.6 for the most massive galaxy in each simulation. Our simulated galaxies agree with the observed relation from a sample of galaxies at $z=6$--8.7 in \citet{mclure.11.sfr} where they connect at $\log(M_{\ast}/\Msun)=8.25$. We also present the growth of galaxy stellar mass and instantaneous star formation rates (SFRs) as a function of cosmic time for these galaxies in the top two panels of Figure \ref{fig:escfrac} (the open symbols represent the time-averaged SFR on 100 Myr time-scales). All these galaxies show significant short-time-scale variabilities in their SFRs, associated with the dynamics of fountains, feedback, and individual star-forming clouds. On larger time-scales (e.g. 100 Myr), the fluctuations in SFRs become weaker and are mostly driven by mergers and global instabilities \citep[see the discussion in][]{hopkins.14.fire}. %Our feedback model regulates the galaxy-averaged star formation efficiency to a low value ($\sim1-2\%$ per dynamical time), in contrast  to many ``sub-grid'' feedback models which tend to form too many stars at high redshifts \citep[e.g.][]{dave.11.galaxy,sparre.14.illustris,fiacconi.15.argo}.

It is worth noticing that our four {\bf z5m10x} simulations have similar global galaxy properties, despite different resolutions and SF prescriptions adopted in these runs. This is because the galaxy-averaged star formation efficiency is regulated by stellar feedback \citep[1--2\% per dynamical time, e.g.][]{hopkins.11.fb,ostriker.11.fb,agertz.13.fb,fg.13.fb}, not the SF criteria \citep{hopkins.13.sf}. However, SF criteria do affect the spatial and density distribution of SF. In {\bf z5m10e}, the SF density threshold $\nc=1~\cm^{-3}$ is comparable or slightly larger than the mean density of the ISM, so that it can be easily reached even in the diffuse ISM. Also, since SF takes place at 100\% efficiency per free-fall time once the gas becomes self-gravitating, many stars form just above the threshold before the gas clouds can further collapse to higher densities. As a consequence, stars are formed either in the diffuse ISM or in gas clouds of densities orders of magnitude lower than those in our standard runs. We emphasize that the {\bf z5m10e} run is not realistic but is designed to mimic star formation models as adopted in low-resolution simulations where the GMCs cannot be resolved.
%Note that the global galaxy properties are very similar between simulations with different star formation density thresholds, but the small-scale ISM structures around individual star particles can be very different between these cases. We emphasize a threshold of $\nc=100~\cm^{-3}$ is more realistic as it corresponds to the typical density inside GMCs.

% Figure 6

\section{Escape Fraction of Ionizing Photons}
\label{sec:escfrac}
We post-process every snapshot with a three-dimensional Monte Carlo radiative transfer (MCRT) code to evaluate the escape fraction of hydrogen ionizing photons from our simulated galaxies. The code is derived from the MCRT code SEDONA base \citep{kasen.06.sedona} and focuses specifically on radiative transfer of hydrogen ionizing photons in galaxies \citep[see][]{fumagalli.11.mcrt,fumagalli.14.mcrt}. For each galaxy, we calculate the intrinsic ionizing photon budget for every star particle within $\Rvir$ to obtain the galaxy ionizing photon production rate $\Qint$. We use the Padova tracks with AGB stars in STARBURST99 with a metallicity $Z=0.0004$ (0.02 $Z_{\odot}$, the closest available to the mean metallicity in our simulations) as our default model (also see Figure~\ref{fig:quanta}). Then we run the MCRT code to compute the rate of ionizing photons that can escape the virial radius $\Qesc$. We define the escape fraction as $\fesc=\Qesc/\Qint$.

% Table 2
\begin{table}
\begin{center}
\caption{Parameters used for the radiative transfer calculation.} 
\label{table:rt}
\begin{tabular}{lcclcc}
\hline\hline
Name & $l_{\rm min}$ & $N_{\rm max}$ & $l_{\rm largest}$ & $N_{\rm photon}$ & $N_{\rm UVB}$ \\
 & (pc) & & (pc) & \\
\hline
{\bf z5m09} & 25 & 250 & $\lesssim40$ & 3e7 & 3e7 \\
{\bf z5m10} & 25 & 300 & $\lesssim80$ & 3e7 & 3e7 \\
{\bf z5m10mr} & 50 & 250 & $\lesssim100$ & 3e7 & 3e7 \\
{\bf z5m10e} & 50 & 300 & $\lesssim80$ & 3e7 & 3e7 \\
{\bf z5m10h} & 50 & 250 & $\lesssim100$ & 3e7 & 3e7 \\
{\bf z5m11} & 50 & 300 & $\lesssim100$ & 4e7 & 4e7 \\
\hline\hline
\multicolumn{5}{l}{{\bf (1)} $l_{\rm min}$: the minimum cell size.} \\
\multicolumn{5}{l}{{\bf (2)} $N_{\rm max}$: the maximum number of cells along each dimension.} \\
\multicolumn{5}{l}{{\bf (3)} $l_{\rm largest}$: the cell size for the largest galaxy in the last snapshot.} \\
\multicolumn{5}{l}{{\bf (4)} $N_{\rm photon}$: number of photon packages being transported.} \\
\hline\hline
\end{tabular}
\end{center}
\end{table}

\subsection{Radiative Transfer Calculation}
\label{sec:mcrt}
We perform the MCRT using a Cartesian grid. We first convert each ``well-resolved'' galaxy identified in our simulations to a cubic Cartesian grid of side length $L$ and with $N$ cells along each dimension. We center the grid at the centre of the galactic halo and choose $L$ equal to two virial radii. The size of a cell $l=L/N$ must be appropriately chosen to ensure convergence. For each simulation, we determine a minimum cell size $l_{\rm min}$ and a maximum $N_{\rm max}$ and then take $N=\min\{L/l_{\rm min},~N_{\rm max}\}$. We have run extensive tests to make sure the parameters $l_{\rm min}$ and $N_{\rm max}$ for each simulation are carefully selected to ensure convergence for every snapshot and maintain reasonable computational expenses. We show examples of convergence tests in Appendix \ref{sec:restest}\footnote{The MCRT calculation converges at much poorer resolution than that we use for hydrodynamics. This is because most of the sources reside in the environment where the ionizing photon optical depth is either $\tau_{\rm UV}\gg1$ or $\tau_{\rm UV}\ll1$. The MCRT calculation will converge as long as the grid captures which limit a star particle is in. However, we emphasize that the high resolution of hydrodynamics is {\it necessary} in order to capture the ISM structure in star-forming regions in the presence of stellar feedback. Low resolution simulations tend to over-predict escape fractions by an order of magnitude (see the discussion in Section \ref{sec:sfc}).}. These parameters are listed in Table \ref{table:rt}. We then calculate the gas density, metallicity, and temperature, at each cell by distributing the mass, internal energy, and metals of every gas particle among a number of cells weighted by their SPH kernel. This conserves mass and energy of gas from the simulation to the grid.

The MCRT method is similar to that described in \cite{fumagalli.14.mcrt}. The radiation field is described by discrete Monte Carlo packets, each representing a large collection of photons of a given wavelength. We emit $N_{\rm star}$ packets isotropically from the  location of the star particles, appropriately sampled by the star UV luminosities. We also emit $N_{\rm UVB}$ packets from the edge of the computational domain in a manner that produces a uniform, isotropic UV background radiation field with intensity given by \citet{fg.09.uvb}. Every photon packet is propagated until it either escapes the grid, or is absorbed somewhere in the grid. Scattering is included in the transport -- i.e., we do not make the on the spot approximation.
 
The photon packets are used to construct estimators of the hydrogen photoionization rates in all cells. The photoionization cross-sections were taken from \citet{verner.96}, the collisional ionization rates from \citet{jefferies.68}, and the radiative recombination rates from \citet{verner.ferland.96}. When calculating the rates, we use the gas temperature from the simulations instead of computing it self-consistently through the radiative transfer\footnote{The simulations take into account many other heating sources (e.g. shocks) besides photoionization. As the radiative transfer code includes collisional ionizations, it is more realistic to take the gas temperature from the simulations than re-computing gas temperature from radiative transfer calculations (in the latter case photoionization would be the only heating source). In regions dominated by photoionization, the uncertainty due to gas temperature is very small, since the recombination rate depends only weakly on temperature.}. We use the case A recombination rates as the transport explicitly treats photon scattering. We assume that 40\% of the metals are in dust phase and adopt a dust opacity of $10^4~\cm^2~{\rm g}^{-1}$ \citep{dwek.98.dust,fumagalli.11.mcrt}. Since the high-redshift galaxies in our simulations tend to be extremely metal-poor, our results do not depend much on dust absorption.

We assume that the gas is in ionization equilibrium, which should be valid for all but the lowest density, highest temperature regions.  Such very low density regions likely do not influence the escape fraction in any case.  We use an iterative method to reach equilibrium, running the MCRT, updating the ionization state of each cell, and then repeating the transport until convergence in the ionization state and escape fraction is reached. We use up to 15 iterations to reach convergence, with typical particle counts per iteration of $3 \times 10^7$ for $N_{\rm star}$ and $N_{\rm UVB}$. We ran tests increasing the particle counts by an order of magnitude to check that the final escape fraction did not change.

% Figure 6
\begin{figure}
\centering
\includegraphics[width=0.5\textwidth]{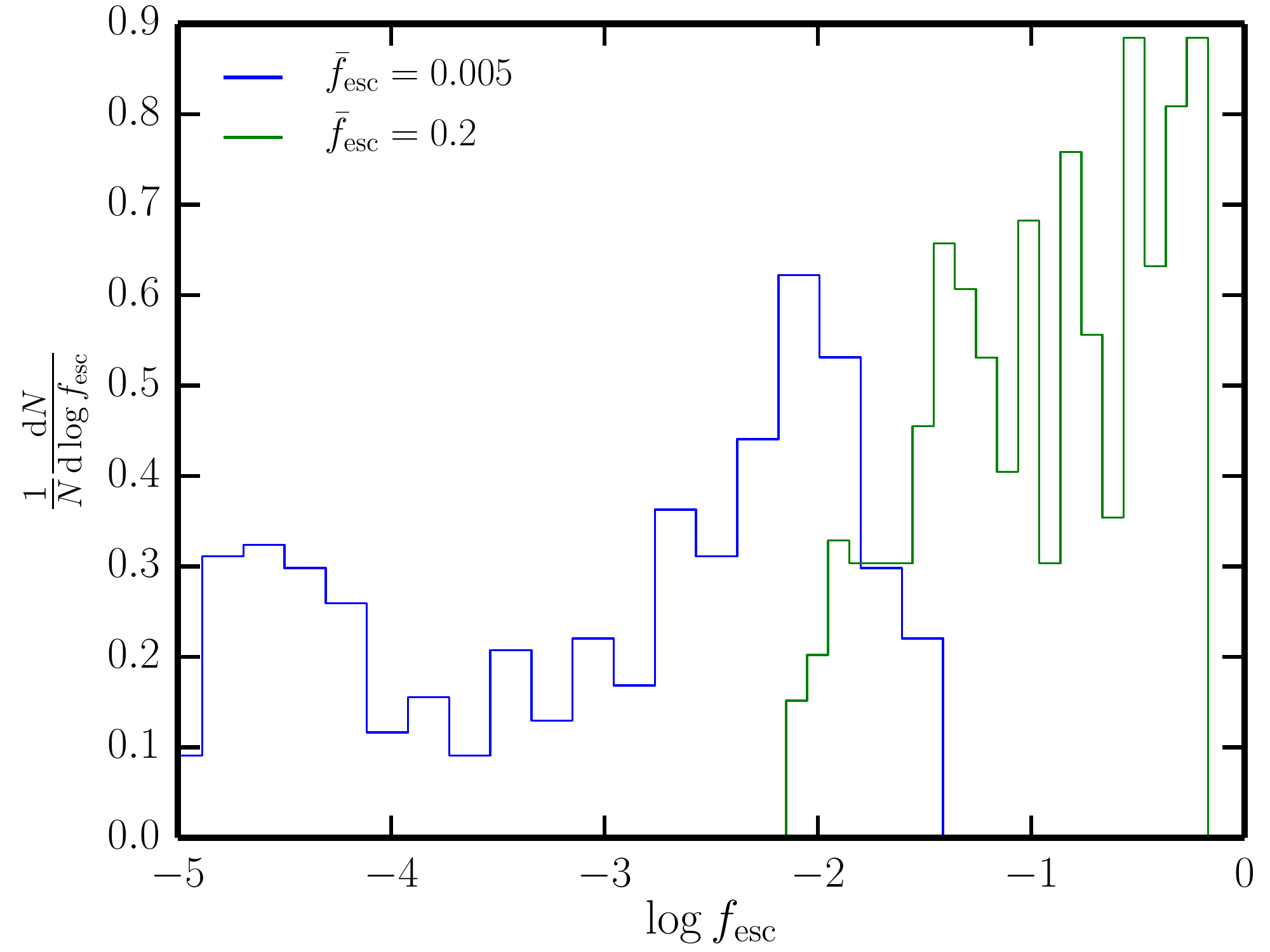}
\caption{Angular distribution of escape fraction for two typical snapshots, with spatially averaged escape fraction 0.005 (blue) and 0.2 (green), respectively. Statistics are obtained from $N=400$ uniformly sampled directions. The broad distribution implies that the ionizing photons that escape to the IGM are highly anisotropic, and that the measured escape fraction from individual galaxies can vary by more than 2 dex depending on the sightline.}
\label{fig:angle}
\end{figure}

% Figure 7
\begin{figure}
\centering
\includegraphics[width=0.5\textwidth]{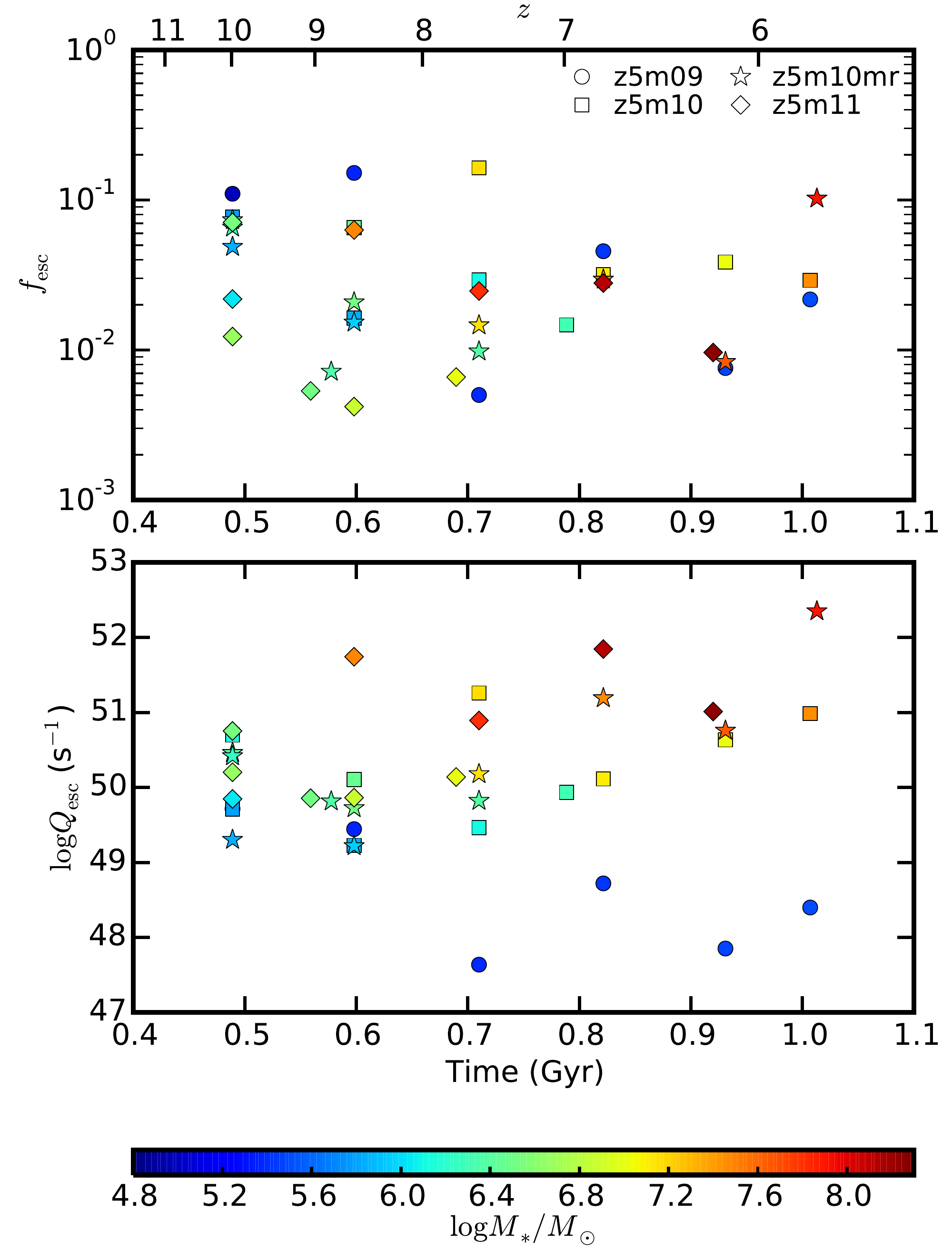}
\caption{Time-averaged escape fraction (top panel) and escaped ionizing photon budget (bottom panel) as a function of cosmic time, color-coded by stellar mass. Different symbols represent the galaxies from different simulations. Points are the escape fraction averaged over 100 Myr ($\Qesc/\Qint$). We see no significant dependence of $\fesc$ on redshift.}
\label{fig:esctime}
\end{figure}

% Figure 8
\begin{figure}
\centering
\includegraphics[width=0.5\textwidth]{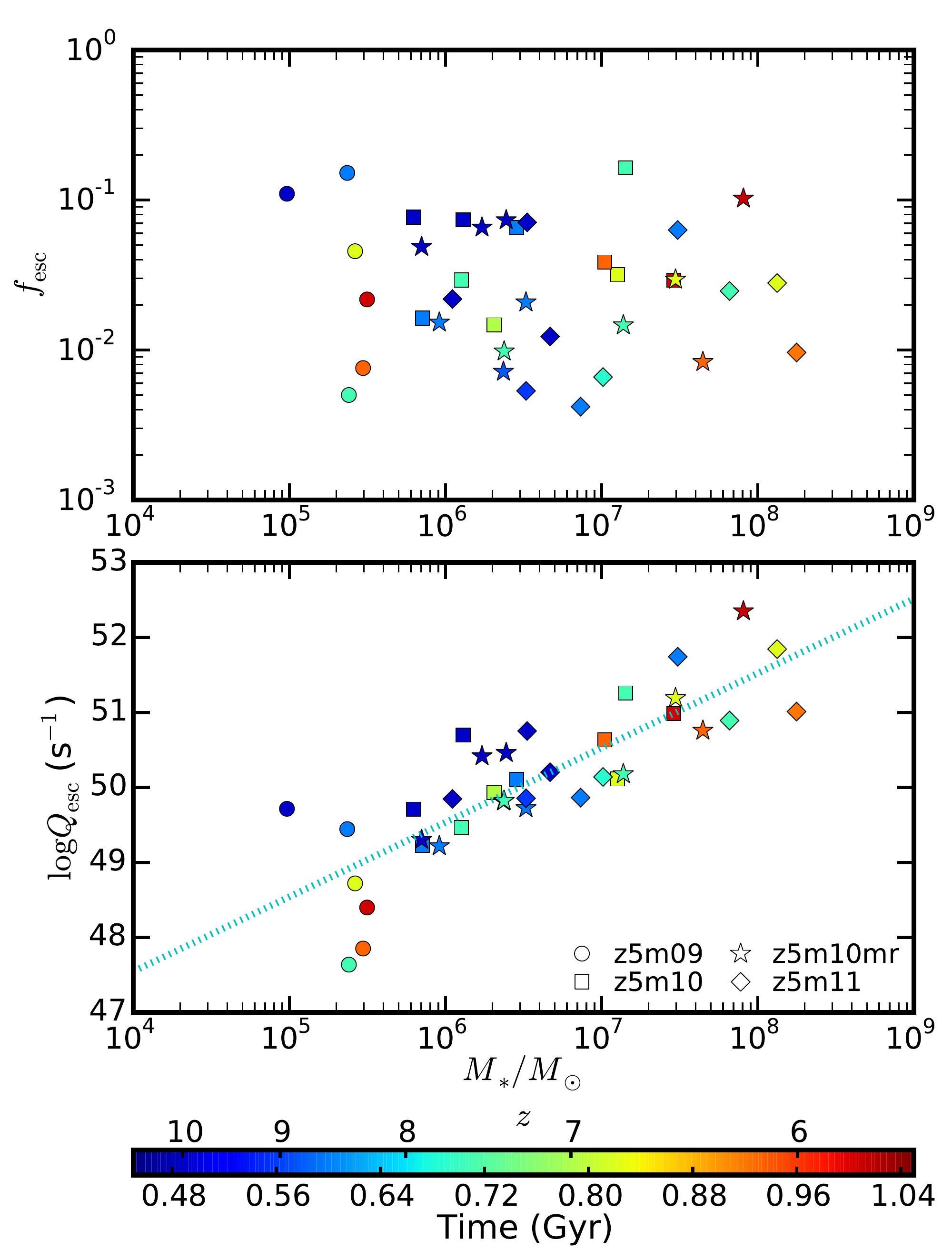}
\caption{Time-averaged escape fraction (top panel) and escaped ionizing photon budget (bottom panel) as a function of stellar mass, color-coded by cosmic time. Different symbols represent the galaxies from different simulations. Points are the escape fraction averaged over 100 Myr. The cyan dotted line in the bottom panel shows the best linear fit of $\log\Qesc=\log(M_{\ast}/\Msun)+43.53$. We see no strong dependence of $\fesc$ on $M_{\ast}$. The dependence of $\Qesc$ on $M_{\ast}$ broadly follows the SFR--$M_{\ast}$ relation.}
\label{fig:escmass}
\end{figure}

\subsection{Instantaneous and Time-averaged Escape Fraction}
In Figure \ref{fig:escfrac}, we present the instantaneous escape fractions ($\fesc$), intrinsic ionizing photon budgets ($\Qint$), and escaped photon budgets ($\Qesc$) as a function of cosmic time for the most massive galaxies in each simulation. We also average $\Qint$ and $\Qesc$ over 100 Myr to obtain the time-averaged escape fractions (the open symbols in Figure \ref{fig:escfrac}). The instantaneous escape fractions show significant time fluctuations, varying between $<0.01\%$ and $>20\%$ from time to time. In our standard runs with default star formation prescriptions, galaxies can reach high escape fractions (10--20\%) only during small amounts of time. For most of the time, the time-averaged escape fractions remain below 5\%. We also calculate the average escape fraction over their entire star formation history (i.e. $z=6$--12). All our standard runs show values between 3--7\%, which confirm low escape fractions on even longer time-scales. The variation in escape fractions on short time-scales is a consequence of feedback and the stochastic formation and disruption of individual star-forming regions, while long time-scale fluctuations are associated with galaxy mergers and intensive starbursts. Note that a high instantaneous escape fraction does not necessarily indicate a high contribution of ionizing photons. For example, although the main galaxy of {\bf z5m11} had an escape fraction around 20\% at $z\sim6.8$, its intrinsic ionizing photon budget $\Qint$ was relatively low at that instant and the time-averaged escape fraction is only $\sim3\%$. Recalling that many models of reionization usually require $\fesc\sim20\%$ if the universe was reionized by galaxies brighter than $\muv>-13$ only \citep[e.g.][]{finkelstein.12.candel,kuhlen.fg.12,robertson.13.udf12}, the escape fractions we find from our simulations are considerably lower than what those models require. 

As long as we properly resolve star formation in dense birth clouds, our results are not sensitive to the details of our star formation prescription\footnote{Previous studies also showed that GMC lifetimes and integrated star formation efficiencies were nearly independent of the instantaneous density threshold and star formation efficiency in dense gas, as long as the clouds were resolved \citep{hopkins.11.fb,hopkins.12.fb}.}. For example, in our {\bf z5m10h} run where we apply $\nc=1000~\cm^{-3}$, the escape fraction is very similar to the standard runs. Also, the similarity between the HR {\bf z5m10} run and the MR {\bf z5m10mr} run shows that our results converge with respect to resolution\footnote{For HR runs where the mass of a star particle is only 10--100 $\Msun$, we also test the effects of the IMF-average approximation. We randomly resample the ionizing photon budgets among individual star particles at a 1:20 ratio according to their ages and repeat the radiative transfer calculations. We find that the escape fractions are very similar. This confirms that the IMF-averaged approximation in HR runs does not affect our results.}. 

However, in our {\bf z5m10e} run where we allow stars form in diffuse gas, the time-averaged escape fraction exceeds 20\% for most of the time. While this toy model results in higher escape fractions, we stress that such a star formation prescription is not consistent with our current understanding of star formation. As such, these predictions are likely not realistic but we include them to illustrate how escape fraction predictions depend sensitively on the ISM model, with our {\bf z5m10e} run being representative of many simulations that do not have sufficient resolution to capture dense ISM structures.

For the fiducial stellar population model we adopt, the majority of the intrinsic ionizing photons are produced by the youngest star particles with age $<3$ Myr (see also Figure \ref{fig:quanta}). These stars are formed in dense, self-gravitating, molecular regions. Most of their ionizing photons are immediately absorbed by their ``birth clouds'' and thus cannot escape the star-forming regions \citep[see also, e.g.][]{kim.13.escape}. When a star particle is older than 3 Myr, a large covering fraction of its birth cloud has been cleared by feedback and thus a significant fraction (order unity) of its ionizing photons are able to propagate to large distances (see e.g. the middle panels of Figure \ref{fig:ism}). Indeed, the ionizing photons that escaped in our simulations mostly come from the star particles of age between 3--10 Myr (also see Section \ref{sec:stellar}). However, the intrinsic ionizing photon budget of a star particle decreases rapidly with age above 3 Myr according to many standard stellar population models. In other words, the escape fractions are primarily determined by small-scale ISM structures surrounding young and intermediate-age star particles. The low escape fractions we find in our simulations are the consequence of the fact that the time-scale for a star particle to clear its birth cloud is comparable to the time-scale for it to exhaust a large amount of its ionizing photon budget. Only when star formation activities are intensive and can last for considerable amount of time, will the ionized regions expand and overlap and thus allow a large fraction of ionizing photons from the youngest stars to escape. For example, the high escape fractions in {\bf z5m10mr} at cosmic time $>1$ Gyr ($z\lesssim6$) are due to the strong and lasting star formation during the past 100 Myr. However, such events are not common in our simulations, since further star formation activity is usually suppressed effectively by stellar feedback.

In Figure \ref{fig:angle}, we show the angular distribution of escape fraction as measured from $N=400$ uniformly sampled directions. We repeat the radiative transfer calculation with ten times more photon packages than listed in Table \ref{table:rt} for two snapshots which have spatially averaged escape fraction $\fesc=0.005$ and 0.2, respectively. The broad distribution of escape factions implies that the ionizing photons escaping to the IGM from galaxies are highly anisotropic. It also indicates that the observationally measured escape fraction from individual galaxies does not necessarily reflect the angle averaged escape fraction from the same object, as it can vary by roughly 2 dex depending on the sightline.

In Figures \ref{fig:esctime} and \ref{fig:escmass}, we compile the time-averaged escape fraction and escaped ionizing photon budget as a function of cosmic time and stellar mass, respectively, for all the ``well-resolved'' galaxies in our standard runs. The symbols are color-coded by stellar mass and cosmic time in Figure \ref{fig:esctime} and \ref{fig:escmass}, respectively. Most of the points lie below $\fesc<5\%$. The escape fraction has a large scatter at fixed cosmic time or stellar mass. We find that there is {\it no} significant dependence of escape fraction on cosmic time or stellar mass. This is consistent with the argument that the escape of ionizing photons is restricted by small-scale ISM structures surrounding the young stellar populations. More simulations are required to study possible redshift and galaxy mass evolution to lower redshifts and over a wider mass interval than sampled by the simulations analyzed in this paper. We also caution that weak trends would be difficult to discern given the time variability found in our simulations and the small size of our simulation sample. The escaped ionizing photon budget depends linearly on stellar mass, with the best fit $\log\Qesc=\log(M_{\ast}/\Msun)+43.53$. This is primarily a consequence of the roughly linear dependence of SFR on stellar mass.

\section{Discussion}
\label{sec:discussion}
We find that instantaneous escape fractions of hydrogen ionizing photons from our simulated galaxies vary between $0.01\%$--$20\%$ from time to time, while time-averaged escape fractions generally remain below 5\%. These numbers are broadly consistent with the wide range of observationally constrained escape fractions measured from variant galaxy samples at $z=0$--3 \citep[e.g.][]{leitet.11,leitet.13,cowie.09,siana.10,bridge.10,iwata.09,boutsia.11,vanzella.12,nestor.13}. 

We obtain much lower escape fractions than previous simulations with ``sub-grid'' ISM, star formation, and feedback models \citep[e.g.][]{razoumov.10,yajima.11}, but our results are more consistent with many recent simulations with state-of-art ISM and feedback models \citep[e.g.][]{kim.13.escape,wise.14,kimm.cen.14,paardekooper.11,paardekooper.15}. Below, we will show that this owes to the failure of the ``sub-grid'' models in resolving stellar birth clouds.

Nevertheless, the escape fractions from our simulated galaxies are still considerably lower than what requires for these galaxies to reionize the universe in many popular models. The tension can be at least partly resolved by invoking galaxies much fainter than what we study in this work, since smaller galaxies have dramatically increasing number densities and possibly much higher escape fractions \citep[e.g.][]{alvarez.12,paardekooper.13,wise.14}. Alternatively, in the rest of this section, we will discuss some physical parameters that might boost the escape fractions in our simulated galaxies. Most of the experiments presented here are for illustrative purposes, but they are worth further exploration in future work in a more systematic and self-consistent way.

\subsection{UV Background}
\label{sec:uvb}
We repeat the radiative transfer calculation for all the snapshots after cosmic time 0.9 Gyr ($z\sim6$) of our {\bf z5m10mr} galaxy with the UV background switched off (the red dotted line in the upper right panel of Figure \ref{fig:escfrac}). The predicted escape fractions does not differ from the previous calculation with the UV background at 0.01\% level, consistent with the results in \citet{yajima.11}. This confirms that the low-density, diffused gas in the galactic halo (which is affected by the UV background) does not affect much the escape of ionizing photons\footnote{However, if the simulations are run without a UV background, gas accretion onto the halo itself can be modified.}.

\subsection{Star Formation Criteria}
\label{sec:sfc}
In our standard simulations, we allow star formation occurs {\it only} in molecular, self-gravitating gas with density above a threshold $\nc=100~\cm^{-3}$. We run {\bf z5m10h} where we adopt $\nc=1000~\cm^{-3}$ for a convergence study. For contrast, we intentionally design {\bf z5m10e} to mimic ``sub-grid'' SF models, where we lower $\nc$ to $1~\cm^{-3}$ and allow extra SF at 2\% efficiency per free-fall time in gas above the threshold but not self-gravitating. In Section \ref{sec:galaxy}, we have confirmed that the global galaxy properties (e.g. star formation rates, stellar masses, UV magnitudes, etc.) are very similar between these runs. However, as it shown in Figure \ref{fig:escfrac}, the escape fraction from {\bf z5m10e} is significantly higher than in other simulations. 

\begin{figure}
\centering
\includegraphics[width=0.5\textwidth]{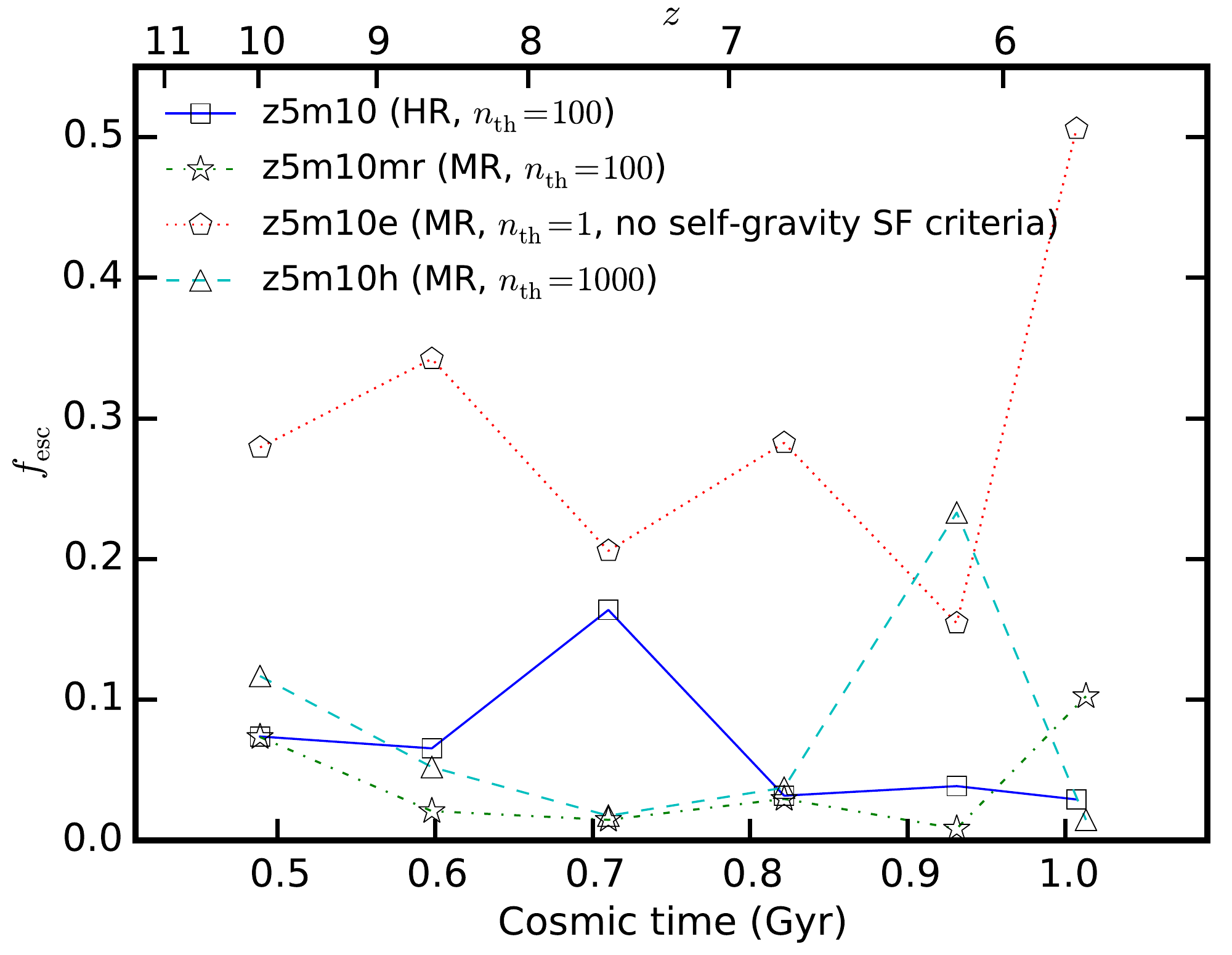}
\caption{Escape fraction with different star formation density prescriptions. The escape fractions averaged over 100 Myr are shown for {\bf z5m10} ($\nc=100~\cm^{-3}$, blue solid), {\bf z5m10mr} ($\nc=100~\cm^{-3}$, green dotted), {\bf z5m10h} ($\nc=1000~\cm^{-3}$, cyan dashed), and {\bf z5m10e} ($\nc=1~\cm^{-3}$, without SF self-gravity criteria, red dotted). For $\nc\gtrsim100~\cm^{-3}$, our results are well-converged with respect to SF density threshold and resolution. However, if SF is allowed in diffuse gas, $\fesc$ can be severely over-estimated.}
\label{fig:density}
\end{figure}

To illustrate this more clearly, we compare the time-averaged (over 100 Myr time-scale) escape fraction of {\bf z5m10}, {\bf z5m10e}, and {\bf z5m10h} in Figure \ref{fig:density}. The qualitative behaviors of escape fraction are very similar between {\bf z5m10}, {\bf z5m10mr}, and {\bf z5m10h}, which further confirm that our results are converged with respect to resolution and SF density threshold (as long as it is much larger than the mean density of the ISM). 

However, in {\bf z5m10e}, the escape fraction is dramatically higher, since many young stars form in the diffuse ISM. Their ionizing photons can then immediately escape the galaxy. We emphasize that the {\bf z5m10e} run is not realistic but mimics ``sub-grid'' SF models as adopted in low-resolution simulations where star formation in dense gas clouds cannot be resolved. This suggests a caution that simulations with ``sub-grid'' SF models can over-predict the escape fraction by an order of magnitude.

% Figure 9

% Figure 10
\begin{figure}
\centering
\includegraphics[width=0.5\textwidth]{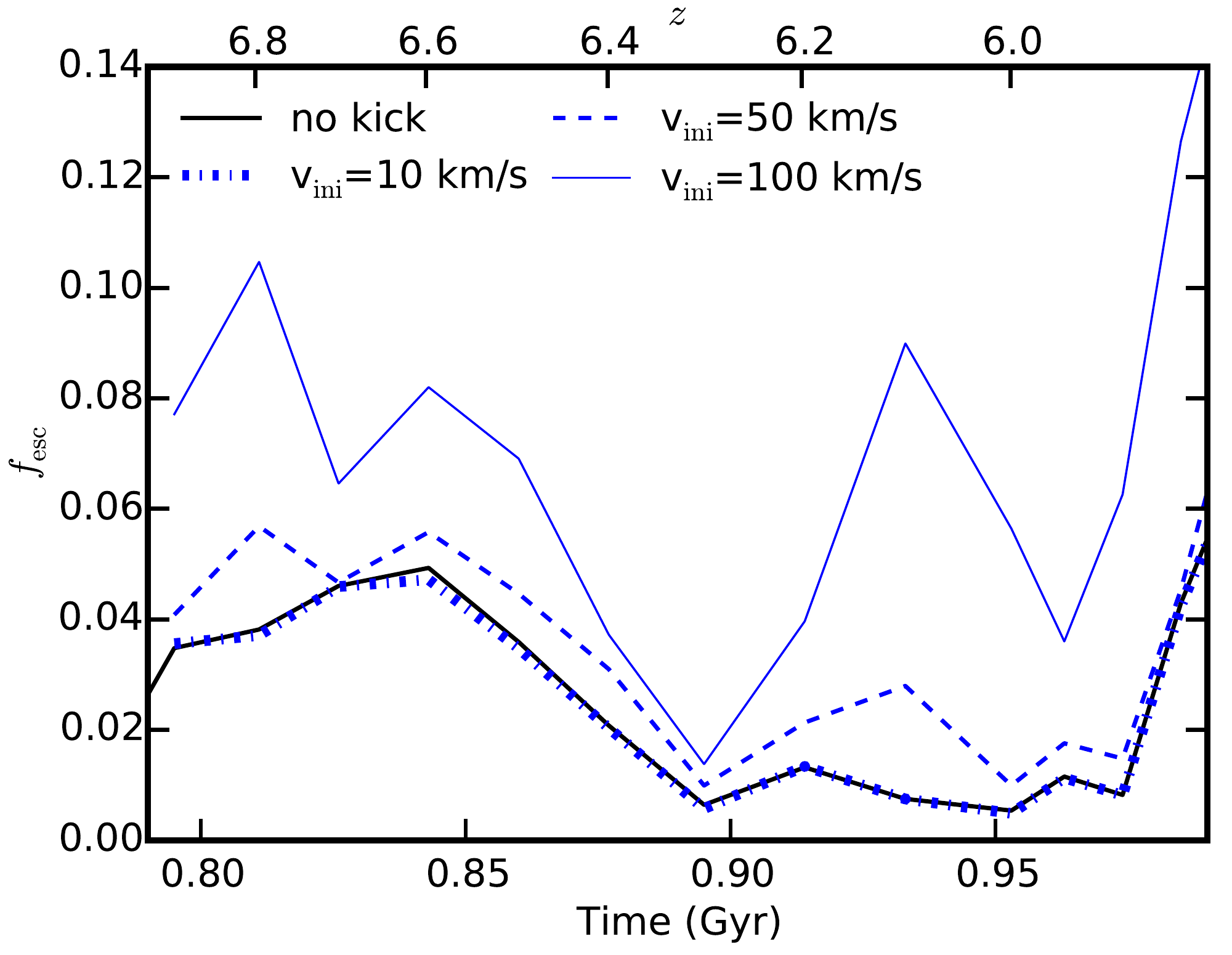}
\caption{Escape fractions in presence of runaway stars. We only show {\bf z5m10mr} during the cosmic time 0.8--1.0 Gyr ($z=6$--7, but the effect in other runs would is similar). Each star particle younger than 10 Myr is kicked from its original position along a random direction with an initial velocity $v_{\rm ini}$. Blue dotted, dashed, and solid lines show the results for $v_{\rm ini}=10 {\rm~km~s^{-1}}$, $50 {\rm~km~s^{-1}}$, and $100 {\rm~km~s^{-1}}$, respectively. The black solid line shows the escape fraction when $v_{\rm ini}=0$ (the same as in Figure~\ref{fig:escfrac}). Typical kick velocities suggested by observations ($\sim 30$ km s$^{-1}$) have only small effects on $\fesc$. Only if the velocities are very large (e.g. $\gtrsim100$ km s$^{-1}$), and an order-unity fraction of stars have been kicked, will this be significant.}
\label{fig:kick}
\end{figure}

% Figure 11
\begin{figure}
\centering
\includegraphics[width=0.5\textwidth]{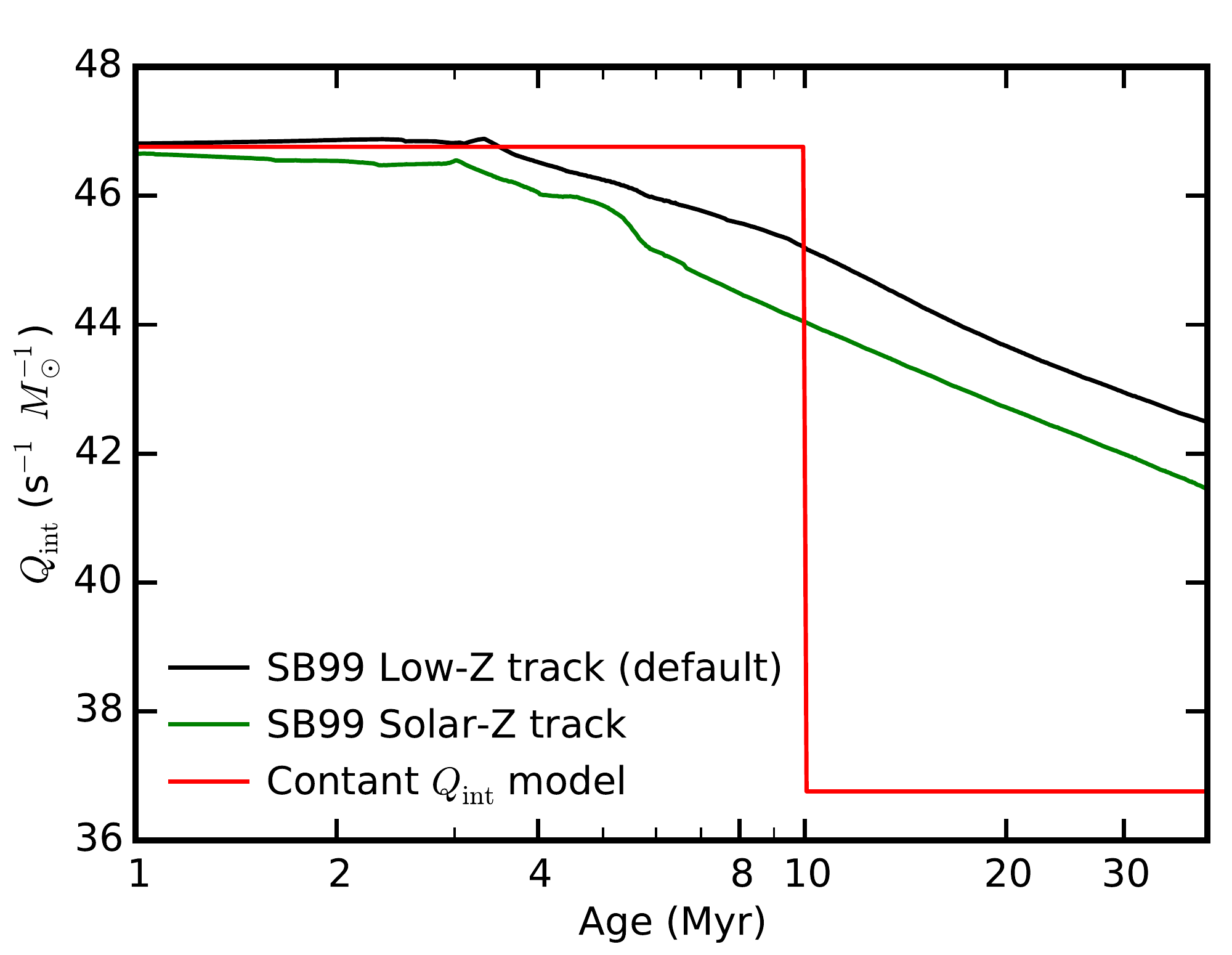}
\caption{Ionizing photon budget per unit mass for a stellar population as a function of its age. The black line shows the STARBURST99 low-metallicity model (our default model). The green line shows the STARBURST99 solar-metallicity model. The red line shows a simple ``constant $Q_{\rm int}$ model'' we consider in Figure \ref{fig:stellar}. This produces a similar number of ionizing photons in the first 3 Myr, but retains the same photon production rate to 10 Myr.}
\label{fig:quanta}
\end{figure}

% Figure 12
\begin{figure}
\centering
\includegraphics[width=0.5\textwidth]{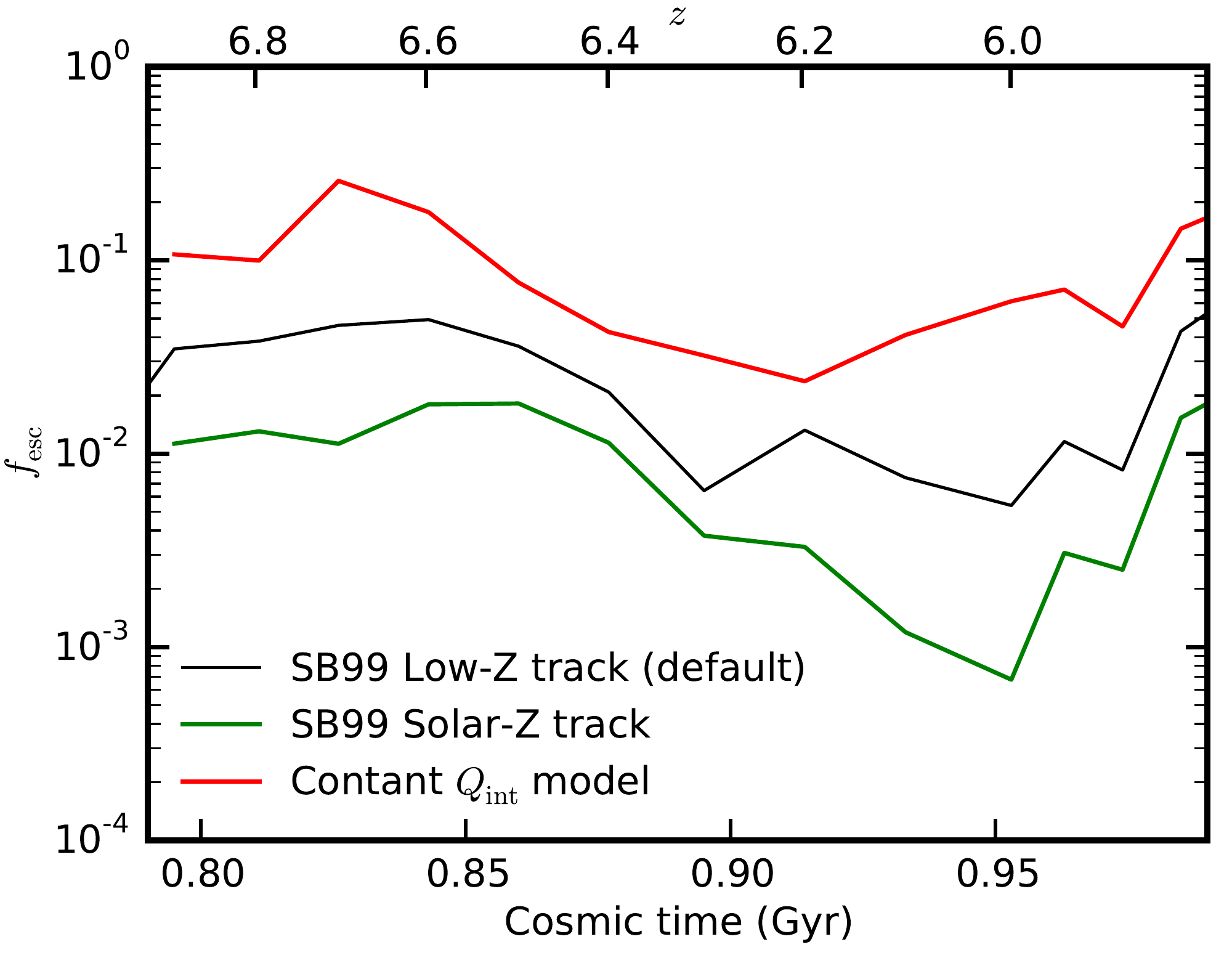}
\caption{Escape fractions calculated by invoking the three different stellar population models shown in Figure \ref{fig:quanta}. We show the results for {\bf z5m10mr} during cosmic time 0.8--1.0 Gyr ($z=6$--7, but the effect in all other runs is similar). The black line shows the results using the STARBURST99 low-metallicity model ($Z=0.0004$, our fiducial model, the same as in Figure~\ref{fig:escfrac}). The green line shows the results using STARBURST99 solar-metallicity model ($Z=0.02$). The red line shows the results when using the ``constant $Q_{\rm int}$ model''. By extending the lifetime of photon production to 3--10 Myr, when the birth clouds have been largely cleared, large $\fesc$ (10--20\%) can be obtained.}
\label{fig:stellar}
\end{figure}

\subsection{Runaway Stars}
There is plenty of evidence that a considerable fraction of O and B stars have high velocities and can travel far from their birth clouds during their lifetime \citep[the ``runaway'' stars, e.g.][]{blaauw.61.runaway,stone.91.runaway,hooger.01.runaway,tetz.11.runaway}. To qualitatively illustrate the effect of these runaway stars on the escape fraction \citep[e.g.][]{conroy.12.runaway}, we move every star particle younger than 3 Myr by a distance $v_{\rm ini} t_{\rm age}$ along a random direction in the snapshots, and repeat the radiative transfer calculation to evaluate the escape fraction as the stars are at their new positions. Here $v_{\rm ini}$ is some initial kicking velocity and $t_{\rm age}$ is the age of the star particle. In principle, it would be more self-consistent if we re-run the whole simulation with runaway star prescription \citep[e.g.][]{kimm.cen.14}. Nonetheless, our simple experiment provides a first estimate of the effects of runaway stars on the escape fraction.

We repeat this experiment for our {\bf z5m10mr} run during cosmic time between 0.8--1.0 Gyr ($z=6$--7) with $v_{\rm ini}=10$, 50, and 100 $\rm km~s^{-1}$, corresponding to a displacement of 30, 150, and 300 pc for a star particle of age 3 Myr. We show the results in Figure \ref{fig:kick}. A small initial velocity of $v_{\rm ini}=10~{\rm km~s}^{-1}$ barely affects the escape fraction since the displacement of a newly formed star particle is $\lesssim30~\pc$, which is much less than the typical size of their birth clouds (see Figure \ref{fig:ism} for an illustration of the ISM structure around young star particles). For $v_{\rm ini}=50~{\rm km~s}^{-1}$, the escape fractions can be boosted by at most 1--2\% (in absolute units, or 20-30\% fractionally). Only for extremely high initial velocity ($\sim100~{\rm km~s}^{-1}$), the escape fractions are enhanced by a few precent, since some young star particles are kicked out of their birth clouds. But these numbers are still somewhat lower than what many reionization models require. Note that observations suggest that only $\sim30\%$ of the OB stars are runaway stars and that the typical velocity of runaway stars is around $30~{\rm km~s}^{-1}$ \citep[e.g.][]{tetz.11.runaway}. Therefore, our experiments suggest that runaway stars will boost the escape fractions by no more than 1\% (in absolute units, or 20\% fractionally)\footnote{One effect not captured in our post-processing experiment which could potentially boost the escape fraction is how feedback from runaway stars would affect the structure of the ISM as they move away from their birth locations. A self-consistent modeling of runaway stars is presented in \citet{kimm.cen.14}. Our result is broadly consistent with theirs for halo masses $\sim10^{10}~\Msun$.}.

%Newly formed stars are always hidden in dense regions. If young stars have some initial velocity when they formed, they might be able to run away from the densest region. We perform an experiment by kicking half of the young star particles by some initial velocity of random direction and let them move a distance equal to the multiply of this velocity and the age of the star particle from its forming place. We then proceed the MCRT calculation as the stars are at their new positions.

\subsection{Stellar Population Models}
\label{sec:stellar}
So far, we have adopted the Padova+AGB stars track of metallicity $Z=0.0004$ (0.02 $Z_{\odot}$) from STARBURST99 model assuming a \citet{kroupa.02.imf} IMF from 0.1--100 $\Msun$ to evaluate the intrinsic ionizing photon budget for each star particle. In this model, the ionizing photon production rate decreases rapidly when the age of a stellar population exceeds 3 Myr. However, there are good reasons to believe that these models suffer from great uncertainties. For example, \citet{steidel.14} emphasized the importance of binary and rotating stars since these stars have high effective temperatures that are required to explain the ionization states of $z=2$--3 star forming galaxies. Moreover, recent theoretical studies suggest that binary star interactions can produce ionizing photons in a stellar population older than 3 Myr; such events are not uncommon \citep[e.g.][]{demink.14.binary}. While these models are very uncertain and still poorly understood, it is not unphysical to invoke more ionizing photons from these populations. To explore the effects of different stellar population models on the escape fractions, we construct a toy model which we refer to the ``constant $\Qint$ model'' to explore its effect on the escape of ionizing photons. In this model, the ionizing photon budget of a stellar population is $5.6\times10^{46}~{\rm s}^{-1}~\Msun^{-1}$ when the population is younger than 10 Myr and suddenly drops to $5.6\times10^{36}~{\rm s}^{-1}~\Msun^{-1}$ when the population is older than 10 Myr. For comparison, we also tabulate the ionizing photon budget using the Padova+AGB stars track of solar metallicity ($Z=0.02$) from STARBURST99 model. We illustrate in Figure \ref{fig:quanta} the intrinsic ionizing photon budget as a function of stellar age for the three models we discuss. Their behaviors are very similar for stellar age $<3$ Myr, after which they start to deviate heavily from each other. The solar-metallicity model has the lowest ionizing photon budget between 3--10 Myr while the constant $\Qint$ model has the highest. 

We repeat the radiative transfer calculation to calculate the escape fraction assuming intrinsic ionizing photon production rate evaluated from these models. In Figure \ref{fig:stellar}, we show the results for our {\bf z5m10mr} run during cosmic time 0.8--1.0 Gyr ($z=6$--7). The escape fractions are very sensitive to the stellar models we use. For the solar-metallicity track model, we get the lowest escape fractions. On the other hand, if we adopt the constant $\Qint$ model, we find the escape fractions are enhanced by almost an order of magnitude. These results further illustrate the picture that the escaped photons come from star particles of age 3--10 Myr, where their birth clouds have been cleared by feedback. Our findings suggest that relatively older stellar populations could contribute a considerable fraction of ionizing photons during reionization, if these populations produce more ionizing photons than what standard stellar population models predict, as motivated by models that include rotation, binaries, and mergers.

\section{Conclusions}
\label{sec:conclusion}
In this work, we present a series of extremely high-resolution (particle mass 20--2000$\Msun$, smoothing length 0.1--4 pc) cosmological zoom-in simulations of galaxy formation down to $z\sim6$, covering galaxy halo masses in $10^9$--$10^{11}~\Msun$, stellar masses in $2\times10^5$--$2\times10^8~\Msun$, and rest-frame ultraviolet magnitudes $\muv=-9$ to $-19$ at that time. This set of simulations include realistic models of the multi-phase ISM, star formation, and stellar feedback (with {\it no} tuned parameters), which allow us to explicitly resolve small-scale ISM structures. Cosmological simulations with these feedback models have been shown to produce reasonable star formation histories, the stellar-mass halo mass relation, the Kennicutt-Schmidt law, the star-forming main sequence, etc., at $z=0$--6 \citep{hopkins.14.fire}. We post-process our simulations with a Monte Carlo radiative transfer code to evaluate the escape fraction of hydrogen ionizing photons from these galaxies. Our main conclusions include the following.

(i) Instantaneous escape fractions have large time variabilities, fluctuating from $<0.01\%$ to $>20\%$ from time to time. In our standard runs, the escape fractions can reach 10--20\% only for a small amount of time. The time-averaged escape fractions (over time-scales 100--1000 Myr) generally remain below 5\%, considerably lower than many recent models of reionization require. 

(ii) As long as star formation is regulated effectively via feedback, the escape fractions are mainly determined by small-scale ISM structures around young and intermediate-age stellar populations. According to standard stellar population models, most of the intrinsic ionizing photons are produced by newly formed star particles younger than 3 Myr. They tend to be embedded in their dense birth clouds, which prevent nearly all of their ionizing photons from escaping. The escaping ionizing photons primarily come from intermediate-age stellar populations between 3--10 Myr, where the dense birth clouds have been largely destroyed by feedback. According to ``standard'' stellar population models, the ionizing photon production rates decline heavily with time at these ages. This leads to the difficulty of getting high escape fractions. 

(iii) The escape fractions do not change if the star formation density threshold increases from 100 to $1000~\cm^{-3}$, as long as stars form in resolved, self-gravitating, dense clouds. On the other hand, if we allow star formation in the diffuse ISM (with some {\it ad hoc} low star formation efficiency), as is adopted in most low-resolution cosmological simulations, the escape fractions can be over-predicted by an order of magnitude. We emphasize that realistic, resolved phase structure of the ISM is critical for converged predictions of escape fractions.

(iv) Applying a fraction of $\sim30\%$ runaway OB stars to our simulations with typical velocity $\sim30~{\rm km~s}^{-1}$ as motivated by many observations can only enhance the escape fraction by at most 1\% (in absolute values, or 20\% fractionally). The effect of runaway stars would not be significant unless a large fraction of the most young stars can obtain dramatically high initial velocity in high-redshift galaxies.

(v) Stellar populations older than 3 Myr may play an important role in reionizing the universe. The escape fractions can be boosted significantly if stellar populations of intermediate ages produce more ionizing photons than what standard stellar population models predict, as suggested by many new stellar population models (e.g. models including rotations, binary interactions, and mergers).

Our simulations are limited in sample size. Also, the simple experiments we present in Section \ref{sec:discussion} do not treat stellar feedback consistently with varying stellar models. Our results motivate further work exploring the effects of IMF variations, stellar evolution models, runaway stars, etc., in a more systematic and self-consistent way. 

%Possible topics ({\it we can re-do RT calculation only for the cheap runs, like {\bf m09}, {\bf m10m}, 
%just in order to make the points}):
%\begin{itemize}
%  \item {\it Runaway OB stars.}
%  \item {\it Old population contribute ionizing photon budget.}
%  \item {\it IMF sampling in high-resolution runs.}
%  \item {\it How UVB affects the galaxy as well as $\fesc$?}
%  \item {\it A larger box at high redshift.} It is worthy doing RT calculation for a high-redshift snapshot, to include the 
%  	whole high-resolution region, with/without UV background. This is to show whether the IGM is ionized by local 
%	star-forming galaxies or UVB (other galaxies or other sources farther away) and how important to treat UVB 
%	and galaxy formation self-consistently.
%  \item {\it RT calculation vs. column density -- if we have the right ionization state in the simulations.} 
%\end{itemize}

%Possible plots:
%\begin{itemize}
%  \item {\it runaway star and old population} A figure of $2\times2$ panels of each point for {\bf m09} 
%  	and {\bf m10m} model, respectively. {\it Probably we could just given the snapshot fluctuation 
%	of $\fesc$.}
%  \item {\it Ionization fraction map.} Using {\bf m09} model at $z=6$, show the map for standard RT 
%  	calculation, direct extracted from simulation, and after IMF re-sampling.
%  \item Similar to the first one, a figure of $2\times2$ panels to compare $\fesc$ fluctuation of LOS 
%  	calculation and IMF re-sampling.
%\end{itemize}

\section*{Acknowledgments}
We thank the anonymous referee for a detailed report and helpful suggestions.
The simulations used in this paper were run on XSEDE computational resources (allocations TG-AST120025, TG-AST130039, and TG-AST140023). The radiative transfer calculations were run on the Caltech compute cluster ``Zwicky'' (NSF MRI award \#PHY-0960291).
D. Kasen is supported in part by a Department of Energy Office of Nuclear Physics Early Career Award, and by the Director, Office of Energy Research, Office of High Energy and Nuclear Physics, Divisions of Nuclear Physics, of the U.S. Department of Energy under Contract No. DE-AC02-05CH11231 and by the NSF through grant AST-1109896. 
Support for PFH was provided by the Gordon and Betty Moore Foundation through Grant 776 to the Caltech Moore Center for Theoretical Cosmology and Physics, by the Alfred P. Sloan Foundation through Sloan Research Fellowship BR2014-022, and by NSF through grant AST-1411920. 
CAFG was supported by NSF through grant AST-1412836, by NASA through grant NNX15AB22G, and by Northwestern University funds. 
D. Kere{\v s} was supported by NSF grant AST-1412153 and funds from the University of California, San Diego. 
EQ was supported by NASA ATP grant 12-APT12-0183, a Simons Investigator award from the Simons Foundation, the David and Lucile Packard Foundation, and the Thomas Alison Schneider Chair in Physics at UC Berkeley.

\bibliography{}

\appendix

% Figure A1
\begin{figure*}
\centering
\includegraphics[width=0.7\textwidth]{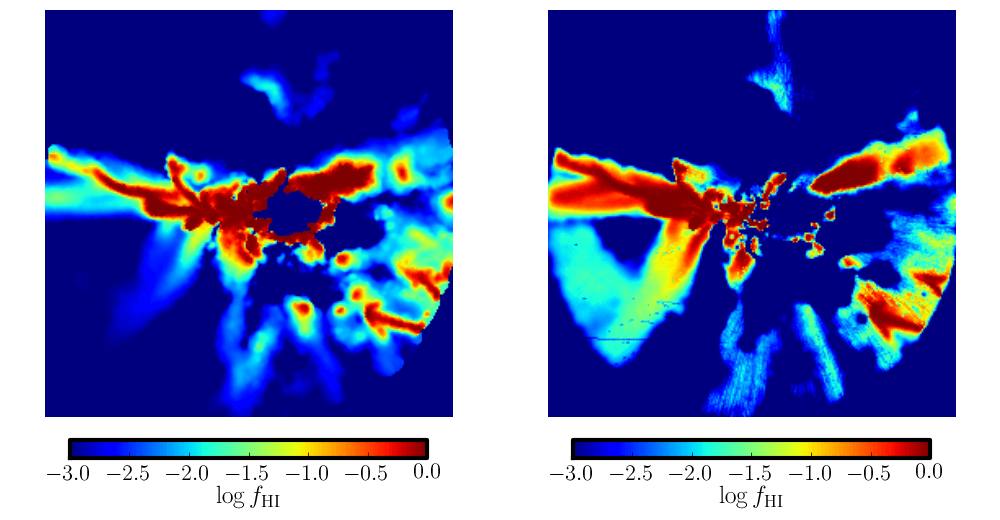}
\caption{Gas neutral fraction on a slice across the galactic centre. We show the snapshot at $z=6$ from {\bf z5m10mr}. {\it Left}: the neutral fraction directly extracted from the simulation. {\it Right}: the neutral fraction recomputed by radiative transfer code. The size of the box equals to two times of the virial radius and there are 250 pixels along each direction. Only regions within the virial radius are shown.}
\label{fig:slice}
\end{figure*}

% Figure A2
\begin{figure*}
\centering
\includegraphics[width=0.7\textwidth]{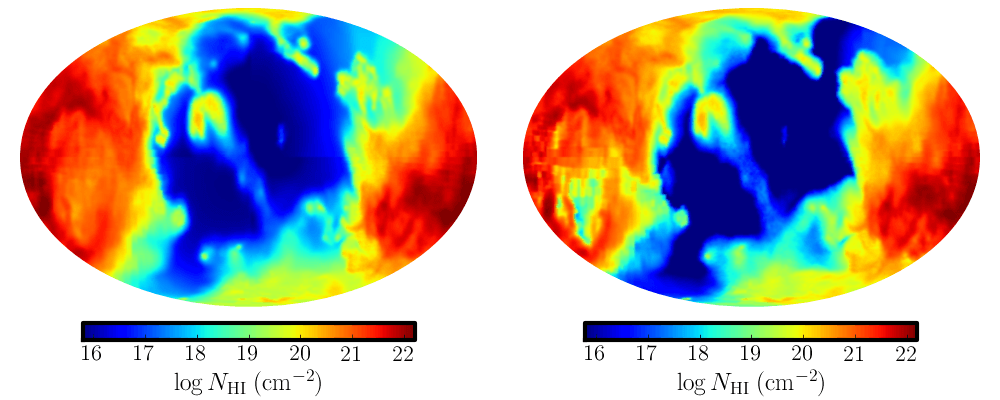}
\caption{Neutral hydrogen column density as viewed from the galactic centre. We show the snapshot at $z=6$ from {\bf z5m10mr}. {\it Left}: the neutral column density directly extracted from the simulation. {\it Right}: the neutral column density recomputed by radiative transfer code.}
\label{fig:map}
\end{figure*}

% Figure A3
\begin{figure}
\centering
\includegraphics[width=0.5\textwidth]{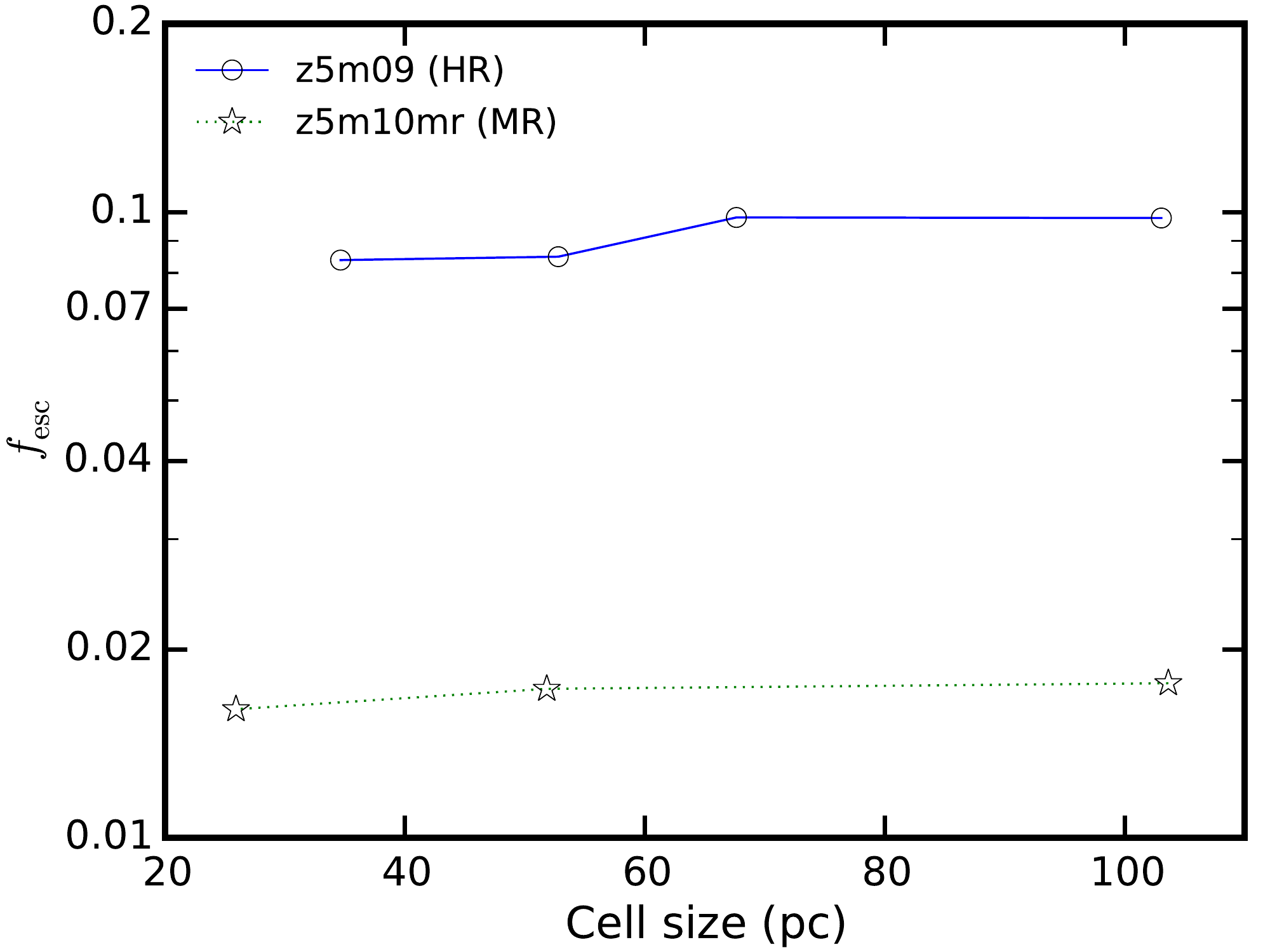}
\caption{Resolution convergence of MCRT calculation. We show two examples, one from {\bf z5m09} (HR) and the other from {\bf z5m10mr} (MR), where we repeat the MCRT calculation for the same galaxy but with different grid cell sizes for the RT calculation. The results are relatively insensitive to the cell size, for cell sizes varying from $l=20$--100 pc.}
\label{fig:restest}
\end{figure}

\section{Comparison of MCRT post-processing to on-the-fly ionization calculations}
\label{sec:appendix}
As described in Section \ref{sec:simulations}, in our simulations, the ionization state of each gas particle at every timestep is determined from the photoionization equilibrium equations described in \citet{katz.96.ionization}, given a uniform and redshift-dependent UV background from \citet{fg.09.uvb} and photon-ionization and photon-electric heating rate from local sources, assuming a local Jeans-length approximation of self-shielding. In the simulations (``on-the-fly''), we model photoionization feedback from star particles in an approximate way -- we move outward from the star particle and ionize each nearest neutral gas particle until the photon budget is completely consumed. In intense star-forming regions, this allows H II regions to expand and overlap and thus approximately captures reasonable ionization states in these regions. However, if the gas distribution is highly asymmetric around an isolated star particle (see, e.g. the middle column in Figure \ref{fig:ism}), the gas ionization states will not be accurately captured. In this work, we follow the propagation of ionizing photons and re-compute the gas ionization state with a Monte Carlo radiative transfer code in post-processing, which will be more accurate in photoionization regions. 

Figure \ref{fig:slice} shows the gas neutral fractions on a slice crossing the galactic centre, of a snapshot at $z=6$ from {\bf z5m10}. Figure \ref{fig:map} shows the neutral hydrogen column density map as viewed from the centre of the galaxy for the same snapshot. In both figures, the left panels are the results before post-processing and the right panels show the results from radiative transfer calculations (using ten times the standard number of photon packages listed in Table \ref{table:rt}). In general, both results agree quite well on large-scale pattern of the neutral gas distribution, although radiative transfer calculations reveal more small structures in star-forming regions. None of our conclusions in this paper are changed qualitatively if we compute the escape fractions using on-the-fly ionization states in the simulations. It is reassuring, both for the present work and for previous studies that used the same approximations, that the approximations used in the simulation code predict ionization structures that are broadly consistent with post-processing radiative transfer calculations.

\section{Resolution convergence for MCRT calculation}
\label{sec:restest}
In Section \ref{sec:mcrt}, we describe that the MCRT calculation is performed on a cubic Cartesian grid of side length $L$ and with $N$ cells along each dimension. In principle, the resolution $l=L/N$ should be small enough to capture the ISM structure, but the number of cells $N^3$ cannot be so big that the calculation is too computationally expensive. After performing extensive convergence tests, we choose $l=25$--100 pc depending on the size of the galactic halo. Here we show two typical examples, one from {\bf z5m09} (HR) and the other from {\bf z5m10mr} (MR), to illustrate the convergence of our MCRT calculation with respect to cell size $l$. As shown in Figure \ref{fig:restest}, we repeat the MCRT calculation for the same galaxy with resolution varying from $l=20$--100 pc and find that the escape fractions do not change appreciably. 

The MCRT calculation converges at much poorer resolution than that we adopt for hydrodynamics. This is because most of the sources reside in an environment where the ionizing photon optical depth is either $\tau_{\rm UV}\gg1$ or $\tau_{\rm UV}\ll1$. In both limits, the MCRT calculation converges even if the exact column density is not captured with great accuracy (e.g. $\tau_{\rm UV}=100$ and $\tau_{\rm UV}=10$ make little difference). However, we emphasize that the high resolution of hydrodynamics is {\it necessary} in order to capture the ISM structure in the presence of star formation and stellar feedback. Low resolution simulations with ``sub-grid'' models tend to over-predict escape fraction (see the discussion in Section \ref{sec:sfc}).

\label{lastpage}

\end{document}